\pdfoutput=1
\PassOptionsToPackage{unicode}{hyperref}
\PassOptionsToPackage{hyphens}{url}
\PassOptionsToPackage{dvipsnames,svgnames,x11names}{xcolor}
\documentclass[
  12pt]{article}

\usepackage{amsmath,amssymb}
\usepackage{iftex}
\ifPDFTeX
  \usepackage[T1]{fontenc}
  \usepackage[utf8]{inputenc}
  \usepackage{textcomp}
\else
  \usepackage{unicode-math}
  \defaultfontfeatures{Scale=MatchLowercase}
  \defaultfontfeatures[\rmfamily]{Ligatures=TeX,Scale=1}
\fi
\usepackage{lmodern}
\ifPDFTeX\else  
\fi
\IfFileExists{upquote.sty}{\usepackage{upquote}}{}
\IfFileExists{microtype.sty}{%
  \usepackage[]{microtype}
  \UseMicrotypeSet[protrusion]{basicmath}
}{}
\makeatletter
\@ifundefined{KOMAClassName}{%
  \IfFileExists{parskip.sty}{%
    \usepackage{parskip}
  }{%
    \setlength{\parindent}{0pt}
    \setlength{\parskip}{6pt plus 2pt minus 1pt}}
}{%
  \KOMAoptions{parskip=half}}
\makeatother
\usepackage{xcolor}
\makeatletter
\ifx\paragraph\undefined\else
  \let\oldparagraph\paragraph
  \renewcommand{\paragraph}{
    \@ifstar
      \xxxParagraphStar
      \xxxParagraphNoStar
  }
  \newcommand{\xxxParagraphStar}[1]{\oldparagraph*{#1}\mbox{}}
  \newcommand{\xxxParagraphNoStar}[1]{\oldparagraph{#1}\mbox{}}
\fi
\ifx\subparagraph\undefined\else
  \let\oldsubparagraph\subparagraph
  \renewcommand{\subparagraph}{
    \@ifstar
      \xxxSubParagraphStar
      \xxxSubParagraphNoStar
  }
  \newcommand{\xxxSubParagraphStar}[1]{\oldsubparagraph*{#1}\mbox{}}
  \newcommand{\xxxSubParagraphNoStar}[1]{\oldsubparagraph{#1}\mbox{}}
\fi
\makeatother

\usepackage{longtable,booktabs,array}
\usepackage{calc}
\usepackage{etoolbox}
\makeatletter
\patchcmd\longtable{\par}{\if@noskipsec\mbox{}\fi\par}{}{}
\makeatother
\IfFileExists{footnotehyper.sty}{\usepackage{footnotehyper}}{\usepackage{footnote}}
\makesavenoteenv{longtable}
\usepackage{graphicx}
\makeatletter
\def\maxwidth{\ifdim\Gin@nat@width>\linewidth\linewidth\else\Gin@nat@width\fi}
\def\maxheight{\ifdim\Gin@nat@height>\textheight\textheight\else\Gin@nat@height\fi}
\makeatother
\setkeys{Gin}{width=\maxwidth,height=\maxheight,keepaspectratio}
\makeatletter
\def\fps@figure{htbp}
\makeatother

\makeatletter
\@ifpackageloaded{caption}{}{\usepackage{caption}}
\AtBeginDocument{%
\ifdefined\contentsname
  \renewcommand*\contentsname{Table of contents}
\else
  \newcommand\contentsname{Table of contents}
\fi
\ifdefined\listfigurename
  \renewcommand*\listfigurename{List of Figures}
\else
  \newcommand\listfigurename{List of Figures}
\fi
\ifdefined\listtablename
  \renewcommand*\listtablename{List of Tables}
\else
  \newcommand\listtablename{List of Tables}
\fi
\ifdefined\figurename
  \renewcommand*\figurename{Figure}
\else
  \newcommand\figurename{Figure}
\fi
\ifdefined\tablename
  \renewcommand*\tablename{Table}
\else
  \newcommand\tablename{Table}
\fi
}
\@ifpackageloaded{float}{}{\usepackage{float}}
\floatstyle{ruled}
\@ifundefined{c@chapter}{\newfloat{codelisting}{h}{lop}}{\newfloat{codelisting}{h}{lop}[chapter]}
\floatname{codelisting}{Listing}

\makeatother
\makeatletter
\@ifpackageloaded{caption}{}{\usepackage{caption}}
\@ifpackageloaded{subcaption}{}{\usepackage{subcaption}}
\makeatother

\ifLuaTeX
  \usepackage{selnolig}
\fi
\usepackage[]{natbib}
\usepackage{bookmark}

\IfFileExists{xurl.sty}{\usepackage{xurl}}{}
\hypersetup{
  pdftitle={A Structural Matrix Autoregression Framework for International Spillovers},
  pdfauthor={Ignacio Moreira Lara; Jan Pr\"user; Christoph Hanck},
  pdfkeywords={Structural vector autoregression; Identification; International spillover; High dimension; Bayesian inference},
  colorlinks=true,
  linkcolor={blue},
  filecolor={Maroon},
  citecolor={Blue},
  urlcolor={Blue},
  pdfcreator={LaTeX via pandoc}}

\usepackage{dsfont}
\usepackage{multirow}
\usepackage{lscape}
\usepackage{tabularx}
\usepackage{threeparttable}
\usepackage{placeins}
\usepackage{algorithm}
\usepackage{algpseudocode}
\AtBeginDocument{}%

\newcommand{\FlexDir}{figures/Flexible}

\newcommand{\ctryblock}[2]{{\overset{\text{#1}}{\begin{pmatrix}\text{GDP}\\ \text{CPI}\end{pmatrix}}}_{\!#2}}
\newcommand{\shockblock}[2]{{\overset{\text{#1}}{\begin{pmatrix}u^{\text{GDP}}\\ u^{\text{CPI}}\end{pmatrix}}}_{\!#2}}
\AtBeginEnvironment{table}{\spacingset{1}}
\AtBeginEnvironment{algorithm}{\spacingset{1}}
\def\spacingset#1{\renewcommand{\baselinestretch}{#1}\small\normalsize}

\begin{document}

\def\spacingset#1{\renewcommand{\baselinestretch}%
{#1}\small\normalsize} \spacingset{1}

\title{\bf A Structural Matrix Autoregression Framework for International Spillovers\thanks{We would like to thank Chenghan Hou, Bowen Fu, Wei Zhang, Boris Blagov, participants from the ESOBE 2026, Statistical Week and the TRR 391 conference for comments. This research was (partially) funded in the course of TRR 391 Spatio-temporal Statistics for the Transition of Energy and Transport (520388526) by the Deutsche Forschungsgemeinschaft (DFG, German Research Foundation). Jan Pr\"user gratefully acknowledges the support of the German Research Foundation (DFG, 468814087). Additional results, extensions and the benchmark comparison are provided in the Online Appendix, available as an ancillary file.}}
\author{Ignacio Moreira Lara\\
  University of Duisburg-Essen and TU Dortmund University
  \and Jan Pr\"user\\
  EBS University of Business and Law
  \and Christoph Hanck\\
  University of Duisburg-Essen}
\date{\today}
\maketitle

\bigskip
\begin{abstract}
Understanding how macroeconomic shocks propagate across countries requires structural models that can jointly identify country-specific shocks and their international transmission. Yet, extending structural vector autoregressions (SVARs) to large multi-country systems is challenging in view of rapidly increasing dimensionality, computational costs, and the proliferation of identifying restrictions. This paper develops a Bayesian Structural Matrix Autoregression (BSMAR) framework that exploits the natural matrix structure of international macroeconomic data. By separating dependence across economic variables from dependence across countries, the framework provides a parsimonious representation that substantially reduces the dimensionality of large structural systems. We develop a Bayesian sampling algorithm for posterior inference that accommodates zero, sign, and ranking (magnitude) restrictions, allowing established SVAR identification schemes to be combined with a novel approach to identifying contemporaneous international spillovers. Applying the model to quarterly supply and demand data for 17 economies, we find evidence of spillovers between countries in both shocks, with the United States as an important transmitter.
\end{abstract}

\noindent%
\vfill

\newpage
\spacingset{1.8}

	\section{Introduction}
	
	Structural vector autoregressions (SVARs) have become a central tool for identifying macroeconomic shocks and studying their dynamic effects on the economy, see, e.g., \citet{sims1980macroeconomics}, \citet{blanchard_quah_1989}, and \citet{kilian2017structural}. Recent applications of sign-restricted SVARs have, in particular, revisited the role of supply and demand disturbances in explaining the post-pandemic inflation episode, see, e.g., \citet{giannone_primiceri_2025}, \citet{bergholt2026inflation}, and \citet{della_chang_jansen_pagliacci_2023}. Yet, despite the inherently international nature of many macroeconomic questions, most structural VAR applications continue to study countries in isolation. This is increasingly difficult to reconcile with the structure of modern economies, where deep trade integration, globally interconnected financial markets and complex production networks allow shocks originating in one country to propagate rapidly across borders. Understanding these spillovers requires structural models that can simultaneously identify country-specific shocks and characterize their transmission across economies.
	
	Extending structural VARs to a large panel of countries, however, presents a fundamental challenge. As the number of countries and variables increases, the dimension of the system and the number of unknown parameters grow rapidly, resulting in severe overparameterization, high computational costs, and a proliferation of identifying restrictions. Consequently, unrestricted multi-country SVARs quickly become infeasible, even for a moderate number of economies given the size of typical balanced macroeconomic datasets.

	To address these challenges this paper develops a Bayesian Structural Matrix Autoregression (BSMAR) approach, together with an efficient sampling algorithm for drawing from its joint posterior distribution. The key insight is that international macroeconomic data naturally have a matrix structure, with one dimension representing economic variables and the other representing countries. Rather than vectorizing the data into a single high-dimensional system, we exploit this two-way structure directly. The resulting framework separates dependence across economic variables from dependence across countries, yielding a parsimonious representation that preserves the multivariate structure of the data while substantially reducing the dimensionality of the model. See the seminal work of \citet{chen2021autoregressive} for the introduction of this class of models.
	
	The matrix formulation resolves the three principal obstacles that have prevented the estimation of large structural international systems. First, it drastically reduces the number of unknown parameters relative to an unrestricted SVAR, mitigating overparameterization and improving statistical efficiency. Second, the resulting parsimonious representation makes Bayesian estimation computationally feasible even in systems containing many countries. Third, and most importantly, it fundamentally simplifies the structural identification problem. Under our framework, the contemporaneous impact matrix admits a Kronecker product decomposition into two variable and country specific components. Rather than identifying one prohibitively large structural impact matrix, identification is carried out separately for economic relationships among variables and for contemporaneous spillovers across countries. This decomposition substantially reduces the number of structural parameters and identifying restrictions while preserving a transparent economic interpretation.
	
	Our Bayesian framework further allows researchers to incorporate a broad class of identifying restrictions, including zero, sign, ranking, and more general nonlinear restrictions. This flexibility makes it possible to combine established SVAR identification schemes for the variable-specific component\footnote{Some examples are sign restriction identification of supply and demand shocks.} with a novel identification framework for the country-specific component. In particular, we use economically motivated restrictions on the country-specific impact matrix to identify structural spillovers across countries. More specifically, a central assumption is that a country-specific shock has its largest contemporaneous impact in the country where it originates. We also allow for soft restrictions that incorporate prior beliefs about the direction and relative magnitude of spillovers without ruling out deviations from these beliefs. 
	
	We demonstrate the empirical relevance of the framework using quarterly data for 17 economies. Motivated by the recent literature on the role of supply and demand shocks in explaining inflation dynamics (e.g., \citet{giannone_primiceri_2025, bergholt2026inflation,della_chang_jansen_pagliacci_2023}) we simultaneously identify 34 country-specific structural shocks, two supply and demand shocks for each economy, and trace the spillovers into other countries. To the best of our knowledge, this is the first study to jointly identify such a large number of country-specific structural shocks. We find substantial evidence of spillovers between countries, with the US acting as an important transmitter. Furthermore, we find that demand shocks emerge as the main driver of the increase in inflation following the COVID-19 pandemic. Lastly, we provide a comparison of the matrix framework with the state-of-the-art Bayesian SVAR with Minnesota shrinkage designed to handle a large number of restrictions \citep{arias_rubio_ramirez_shin_2025}.

	Our paper contributes to two strands of literature. First, we contribute to the rapidly growing literature on matrix-valued time-series models. Matrix autoregressive (MAR) models exploit the two-way structure of matrix-valued data to achieve substantial dimension reduction relative to vectorized VARs. Recent developments include Bayesian estimation, variable selection, nonlinear models, and factor structures, among others, see \citet{celani2024matrix}, \citet{celani2024bayesian}, \citet{bucci2024smooth}, \citet{chan2024bayesian}, \citet{kong2026matrix} and \citet{ye2026multi}. Our contribution in this regard is to develop a structural matrix autoregression that exploits this two-way structure to make high-dimensional structural analysis tractable while allowing researchers to identify contemporaneous structural relationships within countries as well as structural spillovers across countries. 
	
	Second, we contribute to the literature on international macroeconomic spillovers. Existing approaches address the curse of dimensionality by imposing structure on cross-country dependence. For example, dynamic factor models summarize large datasets using a small number of latent factors and are highly effective for studying common business-cycle fluctuations, see, e.g., \citet{forni2001generalized}, \citet{charnavoki2014effects}, \citet{mumtaz2009transmission}, and \citet{stock2005implications}. However, they are not designed to identify a large number of country-specific structural shocks and typically focus on a relatively small number of shocks, see, e.g., \citet{charnavoki2014effects}. Global vector autoregressive (GVAR) models provide another influential framework for studying international macroeconomic linkages, see, e.g., \citet{pesaran2004modeling}, \citet{dees2007exploring}, \citet{crespo2019spillovers}, and \citet{feldkircher2016international}. By connecting country-specific models through predetermined foreign aggregates, GVARs remain tractable for large panels, but at the cost of imposing substantial structure on contemporaneous cross-country interactions. More generally, the existing literature faces a fundamental trade-off: approaches that remain tractable for large panels necessarily impose restrictions on the cross-country dependence structure \citep{camehl2026explains, pfarrhofer2025high}, limiting the identification of country-specific structural spillovers, whereas unrestricted structural models become infeasible as the number of countries and variables increases. Our approach addresses this trade-off by combining the dimensionality reduction of matrix autoregressions with structural identification.
	
	The remainder of the paper is organized as follows. Section~\ref{sec:model} introduces the structural matrix framework, discusses general identification restrictions, and develops the Bayesian sampling algorithm. Section~\ref{sec:empirical} describes the data, presents our empirical identification strategy, and discusses the main empirical findings. The Online Appendix, Section~E, compares our model with a benchmark from the literature. Section~\ref{sec:summary} concludes.
	
	\section{A Large Multi-Country Framework}
	\label{sec:model}
	In this section, we first present the reduced-form Bayesian Matrix Autoregressive (MAR) model. We then formulate its structural representation and discuss the identification of the structural parameters. Finally, we develop a sampling algorithm for drawing from the joint posterior distribution of the elements in the model.
	
	\subsection{The Reduced Form MAR}

	Let $\mathbf{Y}_t$ denote an $n \times k$ matrix of endogenous time series, where $i=1,\ldots,n$ indexes the macroeconomic variables for each country and $j=1,\ldots,k$ indexes the countries, for $t=1,\ldots,T$. That is, the rows of $\mathbf{Y}_t$ correspond to economic variables, while the columns correspond to countries. Hence, the matrix structure reflects the two dimensions that characterize international macroeconomic data: interactions among economic variables and interactions across countries. In total, there are $r=nk$ time series.
	
	The Matrix Autoregressive model, as presented in \citet{chan2025large,samadi2025matrix, chen2021autoregressive}, exploits this structure and is given by
	\begin{equation}
		\label{eq:MAR}
		\mathbf{Y}_t
		=
		\mathbf{A}_1\mathbf{Y}_{t-1}\mathbf{B}_1'
		+\ldots+
		\mathbf{A}_p\mathbf{Y}_{t-p}\mathbf{B}_p'
		+\mathbf{U}_t,
	\end{equation}
	where $\mathbf{A}_1,\ldots,\mathbf{A}_p$ are $n\times n$ coefficient matrices and $\mathbf{B}_1,\ldots,\mathbf{B}_p$ are $k\times k$ coefficient matrices. The matrices $\mathbf{A}_\ell$ capture dynamic interactions among economic variables, whereas $\mathbf{B}_\ell$ capture dynamic interactions across countries. The model therefore allows for a rich but parsimonious representation of the dynamics of the data, making it suitable for high-dimensional settings where the number of variables and countries is jointly large relative to the number of observations.

	The error term is assumed to follow a matrix normal distribution,
	\begin{equation}
		\label{eq:error}
		\mathbf{U}_t
		\sim
		\mathcal{MN}_{n \times k}
		(\mathbf{0},\boldsymbol{\Sigma}_c,\boldsymbol{\Sigma}_r),
	\end{equation}
	where $\boldsymbol{\Sigma}_c$ is a $k\times k$ positive definite column covariance matrix and $\boldsymbol{\Sigma}_r$ is an $n\times n$ positive definite row covariance matrix.\footnote{Note that under the matrix normal distribution, $\operatorname{Cov}(\operatorname{vec}(\mathbf{U}_t))=\boldsymbol{\Sigma}_c\otimes\boldsymbol{\Sigma}_r.$ Thus, the error structure mirrors the two-dimensional organization of the data by separately capturing dependence across countries and across economic variables.} For a more compact representation, define $\mathbf{X}_{t}=\operatorname{blkdiag}(\mathbf{Y}_{t-1},\ldots,\mathbf{Y}_{t-p})$, $
	\mathbf{A}
	=
	(\mathbf{A}_1,\ldots,\mathbf{A}_p)'$ and $
	\mathbf{B}
	=
	(\mathbf{B}_1,\ldots,\mathbf{B}_p)'
	$
	denote the stacked autoregressive coefficient matrices. The MAR model can then be written as
	\begin{equation*}
		\mathbf{Y}_t
		=
		\mathbf{A}'\mathbf{X}_t\mathbf{B}
		+
		\mathbf{U}_t.
	\end{equation*}
	
	To illustrate the MAR form, consider the special case with $p=1$ in which the variable-side matrix is the identity, $\mathbf{A}_1=\mathbf{I}_n$.  In a panel of GDP and CPI series for the United States, Canada and Japan, the block of US equations is given by
	\begin{align*}
		\ctryblock{USA}{t}
		&=
		b_1^{11}\, \ctryblock{USA}{t-1}
		+ b_1^{12}\, \ctryblock{CAN}{t-1}
		+ b_1^{13}\, \ctryblock{JPN}{t-1} 
		+\;  \shockblock{USA}{t},
	\end{align*}
	where $b_1^{j_1j_2}$ denotes the $(j_1,j_2)$ element of $\mathbf{B}_1$, and $u^{\text{GDP}}$ and $u^{\text{CPI}}$ are the reduced-form innovations of the US block. Each country block is thus a linear combination of the corresponding blocks of all countries, its own included, with $b_1^{1j}$ measuring the dynamic propagation from country $j$ to the United States, acting identically on GDP and CPI. When $\mathbf{A}_\ell \neq \mathbf{I}_n$ we have the interactions across variables within a country, since each lagged block is then premultiplied by $\mathbf{A}_\ell$ and GDP and CPI no longer respond proportionally.
	
	A notable characteristic of the MAR model is the equivalent vector representation. Vectorizing yields
	\begin{equation}
		\label{eq:VAR}
		\mathbf{y}_t
		=
		\boldsymbol{\Phi}_1\mathbf{y}_{t-1}
		+\ldots+
		\boldsymbol{\Phi}_p\mathbf{y}_{t-p}
		+\mathbf{u}_t,
	\end{equation}
	where $\mathbf{y}_t=\operatorname{vec}(\mathbf{Y}_t)$, $\mathbf{u}_t=\operatorname{vec}(\mathbf{U}_t)$, $\mathbf{u}_t \sim \mathcal{N}(\mathbf{0}, \boldsymbol{\Sigma}_u)$, 
	$
	\boldsymbol{\Phi}_\ell
	=
	\mathbf{B}_\ell \, \otimes \, \mathbf{A}_\ell,
	$ and $
	\boldsymbol{\Sigma}_u
	=
	\boldsymbol{\Sigma}_c\otimes\boldsymbol{\Sigma}_r.
	$ Hence, the MAR model can be viewed as a VAR in which the large $r\times r$ coefficient matrices and covariance matrix are restricted to have a Kronecker product structure.  Besides providing a familiar framework, the VAR representation of the MAR allows us to summarize our findings with the standard SVAR tools \citep{kilian2017structural}: structural impulse response functions (SIRFs), forecast error variance decompositions (FEVDs) and historical decompositions (HDs). The causal interpretation of these measures depends on the stability condition of the MAR process\footnote{\citet{samadi2025matrix} define stability as the lag polynomial having all roots outside of the unit circle.}. Other interpretations of the MAR model as a GVAR and factor model are presented in \citet{chan2025large}.  
	
	To illustrate the efficiency gains in our model, consider that an unrestricted VAR with $nk$ variables and $p$ lags requires $(nk)^2p$ autoregressive parameters. In contrast, the MAR representation requires only $(n^2+k^2)p$ parameters. For example, with $n=2$ economic variables and $k=17$ countries, an unrestricted VAR requires $1156p$ autoregressive coefficients, compared with only $293p$ in the MAR model. Of course, the reduction becomes increasingly important as the number of countries or variables grows. The MAR model therefore provides a parsimonious yet economically plausible representation that is particularly well suited to high-dimensional settings where estimating an unrestricted VAR would be computationally demanding and statistically inefficient \citep{chen2021autoregressive}.

	\subsection{Structural Matrix Autoregression}
	\label{sec:smar}

	To recover economically meaningful shocks, the reduced-form innovations must be mapped into the vector of structural shocks  $\mathbf{e}_t$:
	\begin{equation}
		\label{eq:SVAR}
		\mathbf{u}_t=\mathbf{B}_0\mathbf{e}_t,
		\qquad
		\mathbf{e}_t\sim\mathcal{N}(\mathbf{0},\mathbf{I}_r),
	\end{equation}
	where $\mathbf{B}_0$ is the structural impact matrix. However, the covariance matrix of the reduced-form innovations only identifies $\mathbf{B}_0$ up to an orthogonal rotation. Specifically, $ \boldsymbol{\Sigma}_u=\mathbf{B}_0\mathbf{B}_0'$, and any orthogonal matrix $\mathbf{Q} \in \mathcal{O}(r)$, for the orthogonal group $\mathcal{O}(r)=\{\mathbf{Q} \in \mathbb{R}^{r\times r}: \mathbf{Q} \mathbf{Q}'=\mathbf{I}_r \} $, yields an observationally equivalent model
	\[
	(\mathbf{B}_0\mathbf{Q}')
	(\mathbf{B}_0\mathbf{Q}')'
	=
	\mathbf{B}_0\mathbf{Q}'\mathbf{Q}\mathbf{B}_0'
	=
	\mathbf{B}_0\mathbf{B}_0'.
	\]
	Therefore, an unrestricted SVAR requires sufficiently many economic restrictions to select a unique rotation among infinitely many observationally equivalent structural representations. This identification problem becomes particularly challenging in large systems. An unrestricted SVAR has $r^2$ structural impact parameters, the number of elements in $\mathbf{B}_0$. Point identification requires at least $r(r-1)/2$ identifying assumptions on $\mathbf{B}_0$. When $n$ or $k$ increases, the number of restrictions required quickly becomes impractical.
	
	To overcome this limitation, we exploit the matrix structure of the VAR and propose a parsimonious representation of the structural impact matrix. Namely we propose the structural bridge:
	\begin{equation}
		\label{eq:BSMAR}
		\mathbf{U}_t =
		\mathbf{B}_r \, \mathbf{E}_t \, \mathbf{B}_c',
		\qquad
		\mathbf{E}_t\sim\mathcal{MN}_{n\times k}(\mathbf{0},\mathbf{I}_k,\mathbf{I}_n),
	\end{equation}
	where $\mathbf{E}_t$ collects the structural shocks in matrix form, so that $\mathbf{e}_t=\operatorname{vec}(\mathbf{E}_t)$. As a by-product, we obtain the structural impact matrix as
	\[
	\mathbf{B}_0=\mathbf{B}_c\otimes\mathbf{B}_r .
	\]
	Hence, this restriction provides two important advantages. First, it substantially reduces the number of unknown structural parameters. Instead of identifying $r^2=n^2k^2$ elements of $\mathbf{B}_0$, the researcher only needs to identify the $n^2+k^2-1$ free elements contained in $\mathbf{B}_r$ and $\mathbf{B}_c$, one element being fixed by the scale normalization discussed below.
	Second, the Kronecker decomposition provides a more transparent economic interpretation of the identifying restrictions. The matrix $\mathbf{B}_r$ captures contemporaneous relationships among variables within a unit and therefore allows researchers to exploit existing identification schemes from the single-country SVAR literature. This is essentially the same separability as for the dynamics, where a single set of $\mathbf{A}_\ell$ matrices governs the variable dynamics of every country and a single set of $\mathbf{B}_\ell$ matrices the cross-country dynamics of every variable. In contrast, $\mathbf{B}_c$ captures cross-unit transmission mechanisms. For $j_1,j_2\in\{1,\ldots,k\}$, we write $b_c^{j_1j_2}$ for the $(j_1,j_2)$ element of $\mathbf{B}_c$. This represents the relative strength of the contemporaneous spillover from country $j_2$ to country $j_1$, and $b_c^{jj}$ is the own effect of country $j$. Consequently, restrictions on $\mathbf{B}_c$ can be directly interpreted as assumptions about economic linkages, such as international spillovers or network effects. Hence, the identification problem is transformed from selecting a rotation of a large unrestricted $r\times r$ matrix $\mathbf{B}_0$ to identifying two lower-dimensional matrices with economically meaningful interpretations. An additional implication of this decomposition is that the covariance structure can be separated into row and column components as $\boldsymbol{\Sigma}_r=\mathbf{B}_r\mathbf{B}_r'$ and $\boldsymbol{\Sigma}_c=\mathbf{B}_c\mathbf{B}_c'$. The price to pay for the gains in efficiency is the rigidity of the model when estimating own country dynamics and spillovers. All interactions are restricted to this bilinear form and the flexibility is constrained.  Whether the Kronecker restrictions are warranted is an empirical question, which we address by formal Bayesian model comparison, see the Online Appendix, Section~E.2.

	An interesting direction for future research is the development of matrix models that relax some separability restrictions while preserving the computational and statistical advantages of the matrix representation. One promising approach is to approximate the single $\left( \mathbf{B}_{\ell} \otimes \mathbf{A}_\ell \right)$ with a sum of Kronecker products, thereby allowing for multiple channels of interaction between countries and indicators while still achieving some dimension reduction. Another possibility is to combine a Kronecker-structured component with an unrestricted component in the structural impact matrix $\mathbf{B}_0$. Such specifications would provide a flexible continuum between the fully separable Matrix SVAR model and an unrestricted SVAR. We believe these extensions could be particularly useful in applications involving a relatively small number of countries, where the additional flexibility can be estimated with reasonable precision. However, as the cross-sectional dimension grows, the number of parameters increases rapidly, leading to a substantial rise in estimation uncertainty and diminishing the practical advantages of these more flexible specifications. We therefore do not apply these weaker restrictions in our application but present both extensions in the Online Appendix, Section~F.

	\subsection{General Structural Identification Restrictions}
	\label{sec:restrictions}
	
	Because the structural-form representation is not identified, additional identifying restrictions are required to narrow the identified set and recover the structural shocks. In this section, we present a general framework for imposing such restrictions within our sampling algorithm presented in Section~\ref{sec:Estimation}. The particular restrictions adopted in the empirical application are described in Section~\ref{sec:emp_id_restrictions}.

	\subsubsection{Parameter Restrictions}
	Define $\boldsymbol{\theta}
	=
	\left(
	\operatorname{vec}(\mathbf{B}_r)',
	\operatorname{vec}(\mathbf{B}_c)'
	\right)'
	\in \mathbb{R}^{d\times 1},$ with $d=n^{2}+k^{2}$, which stacks all structural parameters of the model. Following the notation commonly used in the literature \citep{hou2024large}, we impose linear restrictions on $\boldsymbol{\theta}$ by organizing them into three classes: \emph{zero} restrictions, \emph{sign} restrictions and \emph{relative magnitude} restrictions. Each is a set of linear conditions on $\boldsymbol{\theta}$, obtained by stacking rows using an appropriate selection matrix in $\mathbb{R}^{d}$, whose $i$-th vector we write $\mathbf{q}_i$.
	
	\paragraph{Zero Restrictions.}
	The $r_0$ zero restrictions fix selected elements of $\boldsymbol{\theta}$ at zero. If $\theta_{i_1}=\cdots=\theta_{i_{r_0}}=0$, stack the corresponding basis vectors into
	\begin{equation*}
		\mathbf{R}_{0}
		=
		\begin{bmatrix}
			\mathbf{q}_{i_1} &	\hdots & \mathbf{q}_{i_{r_0}}
		\end{bmatrix}',
		\qquad
		\mathbf{R}_{0}\,\boldsymbol{\theta}=\mathbf{0}.
	\end{equation*}
	This is the special case $\mathbf{d}_i=\mathbf{0}$ of the column-specific equality restriction
	$\mathbf{R}^{E}_i\mathbf{b}_i=\mathbf{d}_i$ considered in \citet{hou2024large}. Such exclusion restrictions rule out specific contemporaneous relationships, for example by setting the contemporaneous effect of one unit on another to zero.

	\paragraph{Sign Restrictions.}
	Sign restrictions fix the direction of selected elements of $\boldsymbol{\theta}$. Let $\psi_1 ,\ldots, \psi_{r_s}$, each in $\{-1,1\}$, be the signs required of $\theta_{i_1}, \ldots, \theta_{i_{r_s}}$, with $+1$ denoting a non-negative element and $-1$ a non-positive one. Stacking the corresponding signed basis vectors for the $r_s$ sign restrictions gives
	\begin{equation*}
		\mathbf{R}_{s}
		=
		\begin{bmatrix}
			\psi_1\mathbf{q}_{i_1} &	\hdots & \psi_{r_s}\mathbf{q}_{i_{r_s}}
		\end{bmatrix}',
		\qquad
		\mathbf{R}_{s}\,\boldsymbol{\theta}\geq\mathbf{0} \text{ elementwise}.
	\end{equation*}
	Typical applications include restricting the direction of a contemporaneous response to a structural shock and/or imposing a sign restriction on cross-unit spillover effects.
	
	\paragraph{Relative Magnitude Restrictions.}
	Similarly the $r_\Delta$ relative magnitude, or ranking, restrictions order two elements of $\boldsymbol{\theta}$ rather than bound either one. Consider the restrictions $\theta_{i_\delta}\geq\theta_{j_\delta}$ for $i_\delta, j_\delta \in \{1, \ldots, r_\Delta \}$\footnote{The same construction yields the two-sided linear inequality restrictions $\mathbf{l}_i<\mathbf{R}^{I}_i\mathbf{b}_i<\mathbf{u}_i$ of \citet{hou2024large}}, and stack the differences of the corresponding basis vectors into
	\begin{equation*}
		\mathbf{R}_{\Delta}
		=
		\begin{bmatrix}
			\mathbf{q}_{i_1}-\mathbf{q}_{j_1} &	\hdots & \mathbf{q}_{i_{r_\Delta}}-\mathbf{q}_{j_{r_\Delta}}
		\end{bmatrix}',
		\qquad
		\mathbf{R}_{\Delta}\,\boldsymbol{\theta}\geq\mathbf{0}.
	\end{equation*}
	This class carries the dominance restriction that we will discuss in Section~\ref{sec:emp_id_restrictions}, which requires a structural coefficient to be at least as large as the competing coefficients in absolute value,
	\[
	b_c^{j_2j_2}\geq |b_c^{j_1j_2}|,
	\qquad
	\forall j_1\neq j_2 .
	\]
	The condition is stated on the own effect including its sign and therefore also implies $b_c^{j_2j_2}\geq0$. Replacing the difference $\mathbf{q}_i-\mathbf{q}_j$ by the pair $\mathbf{q}_i$ and $-\mathbf{q}_i$, and the zero right-hand side by $\underline{\eta}_i$ and $-\overline{\eta}_i$, gives $\underline{\eta}_i\leq\theta_i\leq\overline{\eta}_i$.
	
	Collecting the three classes of restrictions into
	\[
	\mathcal{S}
	=
	\{\mathbf{R}_{0},\,\mathbf{R}_{s},\,\mathbf{R}_{\Delta}\},
	\]
	the identified set is given by \citep[cf.][]{arias_rubio_ramirez_waggoner_2018,read2025fast}
	\begin{align*}
		\mathcal{D}(\mathcal{S})
		&:=
		\left\{
		\boldsymbol{\theta}\in\mathbb{R}^{d}:
		\mathbf{R}_{0}\boldsymbol{\theta}=\mathbf{0},
		\;
		\mathbf{R}\boldsymbol{\theta}\geq\mathbf{0}
		\right\}, \\
		\mathbf{R}
		&=
		\begin{bmatrix}
			\mathbf{R}_{s}' & \mathbf{R}_{\Delta}'
		\end{bmatrix}'.
	\end{align*}
	This set is the $B$-model analogue \citep{kilian2017structural}, of the rotation-based identified set
	\[
	\mathcal{Q}(\boldsymbol{\phi}\mid S)
	=
	\{
	\mathbf{Q}\in\mathcal{O}(r):
	S(\boldsymbol{\phi},\mathbf{Q})\geq\mathbf{0}
	\}
	\]
	defined in \citet{read2025fast}. Because the imposed restrictions generally identify a set rather than a unique structural parameter vector, all elements of $\mathcal{D}(\mathcal{S})$ are observationally equivalent: they generate the same likelihood value and cannot be distinguished using the sample information alone.

	\subsubsection{Soft Parameter Restrictions}
	\label{sec:soft_sign}
	The inequality restrictions above are commonly imposed on the sampler through an accept--reject step \citep{rubio_ramirez_waggoner_zha_2010}. An alternative is proposed by \citet{baumeister2015sign}, who implement sign restrictions via prior information on the structural parameters $\boldsymbol{\theta}$, so that the identifying information is stated on economically meaningful objects (contemporaneous effects and spillovers) rather than on rotations. The conditional posterior is
	\[
	p(\boldsymbol{\theta} \mid \mathbf{Y}, \mathbf{A}, \mathbf{B}) \propto \mathcal{L}(\mathbf{Y} \mid \boldsymbol{\theta}, \mathbf{A}, \mathbf{B})\, p(\boldsymbol{\theta}),
	\]
	with $\mathcal{L}(\cdot)$ the matrix-(Gaussian) likelihood using the stacked observations in $\mathbf{Y},$ and $p(\boldsymbol{\theta})$ the prior. A \emph{soft} sign restriction on $\theta_i$ is a prior that favors a sign but does not enforce a particular sign, e.g., a Gaussian $p(\theta_i) = \mathcal{N}(\theta_i;\, \mu_{\theta_i}, \sigma_{\theta_i}^2)$ with $\mu_{\theta_i} > 0$ and $\sigma_{\theta_i}^2$ chosen so that $P(\theta_i > 0)$ is high (say $0.95$) places most, but not all, of the prior mass on the restricted region. If the likelihood contradicts the restriction sharply enough, the posterior still assigns probability to sign violations, so identification operates through probabilistic weighting of structural models rather than through deterministic exclusion.

	\subsubsection{Normalization Restrictions}
	It is well known that a Kronecker product is unique only up to multiplication by a non-zero scalar $\gamma_\ell$. The likelihood function is unchanged if, for $\ell = 1,\ldots,p$, the pair $(\mathbf{A}_\ell,\mathbf{B}_\ell)$ is replaced by $(\gamma_\ell\, \mathbf{A}_\ell,\gamma_\ell^{-1} \, \mathbf{B}_\ell)$, and likewise if $(\mathbf{B}_c,\mathbf{B}_r)	$ is replaced by $	(\gamma_\ell \, \mathbf{B}_c,\gamma_\ell^{-1}\, \mathbf{B}_r)$.
	We follow a similar approach as \citet{chan2025large} and remove this indeterminacy by fixing the $(1,1)$ element of $\mathbf{B}_\ell$ at one for $\ell = 1, \ldots, p$ and the $(1,1)$ element of $\mathbf{B}_c$ at one. This fixes the scale and determines how the prior mean of these elements is set.

	\subsection{Bayesian Estimation}\label{sec:Estimation}
	In this section we develop the Gibbs sampler to draw from the joint posterior distribution of the elements in the model. First, we introduce the prior distributions. Then, we present the algorithms to sample from the posterior distributions.
	\subsubsection{Prior Specification}
	 To simplify computations we follow \citet{chan2025large} and assume a natural conjugate prior onlPl $\mathbf{A}$ and $\mathbf{B}$. For the unrestricted elements in $\mathbf{B}_r$ and $\mathbf{B}_c$ we assume independent Gaussian distributions.
	\paragraph{Prior Information on $\mathbf{A}$}
	We assume a Gaussian prior on $\mathbf{A}$, $p(\operatorname{vec}(\mathbf{A})\mid\boldsymbol{\Sigma}_r) = \mathcal{N}\!\big(\operatorname{vec}(\underline{\mathbf{A}}), \boldsymbol{\Sigma}_r \otimes \mathbf{V}_{\mathbf{A}}\big)$ with density function:
	\begin{equation}
		p(\mathbf{A})  \propto |\mathbf{V}_{\mathbf{A}}|^{-\frac{n}{2}} |\boldsymbol{\Sigma}_r|^{-\frac{np}{2}} \operatorname{e}^{-\frac{1}{2} \operatorname{tr}\left(\boldsymbol{\Sigma}_r^{-1} (\mathbf{A} - \underline{\mathbf{A}})' \mathbf{V}_\mathbf{A}^{-1} (\mathbf{A} - \underline{\mathbf{A}}) \right)},
	\end{equation}
	where $\underline{\mathbf{A}}$ is the prior mean, $\mathbf{V}_{\mathbf{A}}$ is the $np \times np$ prior covariance matrix, and $\operatorname{tr}(\cdot)$ denotes the trace operator. We specify both to encode Minnesota-type beliefs \citep{litterman1986} adapted to the matrix setting by \citet{chan2025large}. The choice of the prior mean depends on the data specification. In our application, the time series in the data set are in yearly growth rates. Thus, we shrink towards white noise and set $\underline{\mathbf{A}} = \mathbf{0}$. The matrix $\mathbf{V}_{\mathbf{A}}$ is diagonal and is the object through which the shrinkage hyperparameter $\kappa_{\mathbf{A}}$ acts. Recall that $i \in \{1,\dots,n\}$ indexes the rows of $\mathbf{Y}_t$, that is the economic variables, and $j \in \{1,\dots,k\}$ indexes its columns, that is the countries. Let $m_A=(\ell-1)n+i$ denote the row of $\mathbf{A}$ that carries the coefficients on row variable $i$ at lag $\ell \in \{1,\dots,p\}$. To fix the scale, let $\hat{s}^2_{i,j}$ be the residual variance of an AR(4) regression fitted to the single series $y_{t,ij}$, and let $\hat{s}^2_{i,\bullet} \;=\; \frac{1}{k}\sum_{j=1}^{k} \hat{s}^2_{i,j}$
	denote its average across the $k$ countries in row $i$. The corresponding diagonal element of $\mathbf{V}_{\mathbf{A}}$ then is
	\begin{equation}
		v_{\mathbf{A},m_A m_A} \;=\; \frac{\kappa_{\mathbf{A}}}{\ell^{2}\, \hat{s}^{2}_{i,\bullet}},
		\qquad m_A = (\ell-1)n + i.
		\label{eq:VA}
	\end{equation}
	The prior variance thus factors into three pieces. The term $\ell^{-2}$ shrinks higher-order lags towards zero more aggressively, reflecting the belief that distant lags carry little information. The term $\hat{s}^{-2}_{i,\bullet}$ makes the prior invariant to the units in which each row of $\mathbf{Y}_t$ is measured. Finally, $\kappa_{\mathbf{A}}$ scales the whole matrix and therefore governs the \emph{overall} shrinkage strength: as $\kappa_{\mathbf{A}} \to 0$ the prior collapses onto $\underline{\mathbf{A}}$, while a large $\kappa_{\mathbf{A}}$ leaves $\mathbf{A}$ essentially unrestricted. Equivalently, $\mathbf{V}_{\mathbf{A}} = \kappa_{\mathbf{A}} \mathbf{C}_{\mathbf{A}}$, where $\mathbf{C}_{\mathbf{A}}$ is a (fixed) diagonal matrix determined by the data and the lag structure alone.

	Rather than fixing $\kappa_{\mathbf{A}}$ at some default value, we treat it as unknown and let the data determine its value, following \citet{giannone_lenza_primiceri_2015}. We place a hierarchical gamma prior $\kappa_{\mathbf{A}} \sim \mathcal{G}(c_{\mathbf{A},1}, c_{\mathbf{A},2})$ and set $c_{\mathbf{A},1} = c_{\mathbf{A},2} = 5$, a weakly informative choice centered at one.
	\paragraph{Prior Information for $\mathbf{B}$ }
	The prior on $\mathbf{B}$ is close in spirit, the density function given by:
	\begin{equation}
		p(\mathbf{B})  \propto |\mathbf{V}_{\mathbf{B}}|^{-\frac{k}{2}} |\boldsymbol{\Sigma}_c|^{-\frac{kp}{2}} \operatorname{e}^{-\frac{1}{2} \operatorname{tr}\left(\boldsymbol{\Sigma}_c^{-1} (\mathbf{B} - \underline{\mathbf{B}})' \mathbf{V}_\mathbf{B}^{-1} (\mathbf{B} - \underline{\mathbf{B}}) \right)}.
	\end{equation}
	We set the prior mean as $\underline{\mathbf{B}} = (\mathbf{I}_k, \ldots, \mathbf{I}_k)'$ to represent no spillovers and country-homogeneous beliefs on $\mathbf{B}$. The matrix $\mathbf{V}_{\mathbf{B}}$ is built symmetrically to \eqref{eq:VA}: let $m_B=(\ell-1)k+j$ denote the row of $\mathbf{B}$ that carries the coefficients on column country $j$ at lag $\ell$, and let $\hat{s}^{2}_{\bullet,j} = n^{-1}\sum_{i=1}^{n} \hat{s}^{2}_{i,j}$ denote the average AR(4) residual variance down column $j$. We set
	\[
	v_{\mathbf{B},m_B m_B} \;=\; \frac{\kappa_{\mathbf{B}}}{\ell^{2}\, \hat{s}^{2}_{\bullet,j}},
	\qquad m_B = (\ell-1)k + j ,
	\]
	so that $\kappa_{\mathbf{B}}$ determines how informative the prior is: a small $\kappa_{\mathbf{B}}$ pins $\mathbf{B}_\ell$ at the identity, a large one results in an uninformative prior, and again $\mathbf{V}_{\mathbf{B}} = \kappa_{\mathbf{B}} \mathbf{C}_{\mathbf{B}}$ with $\mathbf{C}_{\mathbf{B}}$ fixed and diagonal. We use a hierarchical gamma prior $\kappa_{\mathbf{B}} \sim \mathcal{G}(c_{\mathbf{B},1}, c_{\mathbf{B},2})$ with $c_{\mathbf{B},1} = c_{\mathbf{B},2} = 5$ and estimate it from the data.

	\paragraph{Priors on $\mathbf{B}_r$ and $\mathbf{B}_c$}
	We specify \emph{Gaussian} priors on the free elements of $\mathbf{B}_r$ and $\mathbf{B}_c$,
	\[
	\theta_i \sim \mathcal{N}\!\left(\mu_{\theta_i},\, \sigma^2_{\theta_i}\right), \qquad \theta_i \in \boldsymbol{\theta},
	\]
	with $\boldsymbol{\mu}_{\theta}$ and the diagonal $\boldsymbol{\Sigma}_{\theta} = \operatorname{diag} (\sigma^2_{\theta_i})$ collecting the prior means and variances discussed in Section~\ref{sec:soft_sign}. The prior on the covariance blocks is therefore \emph{induced} by the prior on the structural parameters rather than elicited separately, which is what allows the identifying restrictions to be stated on economically interpretable objects. 
	
	\subsubsection{Conditional Posteriors}
	\paragraph{Sampling $\mathbf{A}$ and $\mathbf{B}$.}
	
	Conditional on $\mathbf{B}_r$ and $\mathbf{B}_c$, the two covariance matrices are known, and the coefficient blocks retain their conjugate form. The conditional posteriors for $\mathbf{A}$ and $\mathbf{B}$ are 
	\begin{equation}
		\operatorname{vec}(\mathbf{A}) \mid \mathbf{Y}, \boldsymbol{\Sigma}_r, \mathbf{B}, \boldsymbol{\Sigma}_c, \kappa_{\mathbf{A}}, \kappa_{\mathbf{B}} \sim \mathcal{N}\!\big(\operatorname{vec}(\widehat{\mathbf{A}}), \boldsymbol{\Sigma}_r \otimes \mathbf{K}_{\mathbf{A}}^{-1}\big),
	\end{equation}
	with
	\begin{equation*}
		\mathbf{K}_\mathbf{A}  
		= \mathbf{V}_\mathbf{A}^{-1} 
		+ \sum_{t=1}^T\mathbf{X}_t \mathbf{B} \boldsymbol{\Sigma}_c^{-1} \mathbf{B}' \mathbf{X}_t',
		\qquad
		\widehat{\mathbf{A}}
		= \mathbf{K}_{\mathbf{A}}^{-1} \left( \mathbf{V}_{\mathbf{A}}^{-1} \underline{\mathbf{A}}
		+ \sum_{t=1}^T \, \mathbf{X}_t \mathbf{B} \boldsymbol{\Sigma}_c^{-1} \mathbf{Y}_t' \right).
	\end{equation*}
	
	Similarly, the conditional posterior of $\mathbf{B}$ is given by
	\begin{equation}
		\operatorname{vec}(\mathbf{B}) \mid \mathbf{Y}, \boldsymbol{\Sigma}_c, \mathbf{A}, \boldsymbol{\Sigma}_r, \kappa_{\mathbf{A}}, \kappa_{\mathbf{B}} \sim \mathcal{N}\!\big(\operatorname{vec}(\widehat{\mathbf{B}}), \boldsymbol{\Sigma}_c \otimes \mathbf{K}_{\mathbf{B}}^{-1}\big),
	\end{equation}
	where
	\begin{equation*}
		\mathbf{K}_\mathbf{B} 
		= \mathbf{V}_\mathbf{B}^{-1} 
		+ \sum_{t=1}^T \, \mathbf{X}_t' \mathbf{A} \boldsymbol{\Sigma}_r^{-1} \mathbf{A}' \mathbf{X}_t,
		\qquad
		\widehat{\mathbf{B}} 
		= \mathbf{K}_\mathbf{B}^{-1} \left( \mathbf{V}_\mathbf{B}^{-1} \underline{\mathbf{B}}
		+ \sum_{t=1}^T \, \mathbf{X}_t' \mathbf{A} \boldsymbol{\Sigma}_r^{-1} \mathbf{Y}_t \right).
	\end{equation*}
	
	One of the key advantages of this approach is its computational efficiency. In particular, drawing the autoregressive component of the model costs $\mathcal{O}(n^3p^3)$ for $\mathbf{A}$ and $\mathcal{O}(k^3p^3)$ for $\mathbf{B}$, rather than the $\mathcal{O}\big((nk)^3p^3\big)$ required by a corresponding unrestricted VAR; see \citet{chan2025large} for a detailed discussion. For the normalization restrictions we implement the algorithm by \citet{rue_held_2005}.
	
	\paragraph{Sampling $\kappa_{\mathbf{A}}$ and $\kappa_{\mathbf{B}}$.}
	Because the shrinkage parameters enter the prior variances multiplicatively through $\mathbf{V}_{\mathbf{A}} = \kappa_{\mathbf{A}}\mathbf{C}_{\mathbf{A}}$ and $\mathbf{V}_{\mathbf{B}} = \kappa_{\mathbf{B}}\mathbf{C}_{\mathbf{B}}$, combining the gamma priors with the coefficient priors yields generalized inverse Gaussian (GIG) conditional posteriors,
	\begin{align*}
		\kappa_{\mathbf{A}} \mid \mathbf{A}, \boldsymbol{\Sigma}_r
		&\sim
		\mathcal{GIG}\!\left(
		c_{\mathbf{A},1}-\tfrac{n^{2}p}{2},\;
		2c_{\mathbf{A},2},\;
		\operatorname{tr}\!\big(\boldsymbol{\Sigma}_r^{-1}(\mathbf{A}-\underline{\mathbf{A}})'\mathbf{C}_{\mathbf{A}}^{-1}(\mathbf{A}-\underline{\mathbf{A}})\big)
		\right), \\
		\kappa_{\mathbf{B}} \mid \mathbf{B}, \boldsymbol{\Sigma}_c
		&\sim
		\mathcal{GIG}\!\left(
		c_{\mathbf{B},1}-\tfrac{k^{2}p}{2},\;
		2c_{\mathbf{B},2},\;
		\operatorname{tr}\!\big(\boldsymbol{\Sigma}_c^{-1}(\mathbf{B}-\underline{\mathbf{B}})'\mathbf{C}_{\mathbf{B}}^{-1}(\mathbf{B}-\underline{\mathbf{B}})\big)
		\right).
	\end{align*}
	To draw from this density we use the algorithm by \citet{devroye2014random}. The covariance matrices are then rebuilt from the new draws.
	
	\paragraph{Sampling $\mathbf{B}_r$ and $\mathbf{B}_c$.}
	The structural factors do not inherit the conjugacy of the coefficient blocks. Conditional on $(\mathbf{A}, \mathbf{B})$ the reduced-form innovations are known,
	\[
	\mathbf{U}_t(\mathbf{A},\mathbf{B}) \;=\; \mathbf{Y}_t - \sum_{\ell=1}^{p} \mathbf{A}_\ell \mathbf{Y}_{t-\ell} \mathbf{B}_\ell' ,
	\qquad
	\mathbf{u}_t \;=\; \operatorname{vec}\big(\mathbf{U}_t\big) \;=\; \mathbf{B}_0\, \mathbf{e}_t,
	\qquad
	\mathbf{B}_0 = \mathbf{B}_c \otimes \mathbf{B}_r ,
	\]
	with $\mathbf{e}_t = \operatorname{vec}(\mathbf{E}_t) \sim \mathcal{N}(\mathbf{0}, \mathbf{I}_{r})$. Gathering the free elements of $\mathbf{B}_r$ and $\mathbf{B}_c$ in the vector $\boldsymbol{\theta}$, the conditional log-likelihood is
	\begin{equation}
		\log \mathcal{L}(\mathbf{Y} \mid \boldsymbol{\theta}, \mathbf{A}, \mathbf{B})
		= \alpha
		- T\Big( k \log\big|\det \mathbf{B}_r\big| + n \log\big|\det \mathbf{B}_c\big| \Big)
		- \frac{1}{2} \sum_{t=1}^{T} \operatorname{tr}\!\left( \boldsymbol{\Sigma}_r^{-1} \mathbf{U}_t \boldsymbol{\Sigma}_c^{-1} \mathbf{U}_t' \right),
		\label{eq:lik_theta}
	\end{equation}
	where the log-determinant term uses $\log|\det \mathbf{B}_0| = k \log|\det \mathbf{B}_r| + n \log|\det \mathbf{B}_c|$, the Jacobian of the map from shocks to innovations. Combining \eqref{eq:lik_theta} with the Gaussian prior $\boldsymbol{\theta} \sim \mathcal{N}(\boldsymbol{\mu}_{\theta}, \boldsymbol{\Sigma}_{\theta})$ of the previous subsection and the identified set $\mathcal{D}(\mathcal{S})$ gives the conditional posterior
	\begin{equation}
		p(\boldsymbol{\theta} \mid \mathbf{Y}, \mathbf{A}, \mathbf{B})
		\;\propto\;
		\mathcal{L}(\mathbf{Y} \mid \boldsymbol{\theta}, \mathbf{A}, \mathbf{B})\;
		p(\boldsymbol{\theta} \mid \boldsymbol{\mu}_{\theta}, \boldsymbol{\Sigma}_{\theta})\;
		\mathbf{1}\left\{ \boldsymbol{\theta} \in \mathcal{D}(\mathcal{S}) \right\}.
		\label{eq:post_theta}
	\end{equation}
	Because $\boldsymbol{\theta}$ enters the likelihood through both the determinant and the inverse of $\mathbf{B}_0$, \eqref{eq:post_theta} belongs to no known family: there is no closed-form draw and no conjugate update. We sample it with an elliptical slice sampler, which we describe next.
	\subsubsection{Elliptical Slice Sampling (ESS)}
	\label{sec:ess}
	
	To sample from the conditional posterior \eqref{eq:post_theta} of the structural parameters we implement the elliptical slice sampling algorithm of \citet{murray2010elliptical}. The method is tailored to target a posterior constructed from a Gaussian prior multiplied by an arbitrary likelihood and it requires neither a Metropolis accept-reject step nor any stepsize tuning.\footnote{ We have also experimented with the DREAM algorithm by \citet{Vrugt2016_DREAM}, an efficient Metropolis--Hastings scheme that, unlike the elliptical slice sampler, accommodates arbitrary (non-Gaussian) priors on the structural parameters.}
	
	The closest framework to ours is by \citet{arias_rubio_ramirez_shin_2025}, who embed the elliptical slice sampler in a Gibbs sampler for sign-identified SVARs and show that it remains fast in large dimensions where accept-reject sampling becomes infeasible. Our sampler shares that logic and differs in what the slice sampler moves. They work in the orthogonal reduced-form parameterization of \citet{arias_rubio_ramirez_waggoner_2018} and update each block in turn. The proposals from their algorithm come from the posterior distributions and satisfy the restrictions needed by the orthogonal matrices. We instead parameterize the structural impact matrix directly through $\boldsymbol{\theta}$ and move its free elements, so that the ellipse carries the Gaussian prior $\mathcal{N}(\boldsymbol{\mu}_{\theta}, \boldsymbol{\Sigma}_{\theta})$ and the likelihood \eqref{eq:lik_theta} enters the slice threshold. In our setting the restrictions are enforced as a feasibility check that shrinks the bracket, while their algorithm only accepts draws where the orthogonal matrix fulfills the conditions.

	\paragraph{The Proposal.} Let $\boldsymbol{\theta}^{(m-1)}$ denote the current state. The sampler draws an auxiliary direction from the prior, $\boldsymbol{\nu}\sim \mathcal{N}(\mathbf{0},\boldsymbol{\Sigma}_{\theta})$, with candidate draws along the ellipse through $\boldsymbol{\theta}^{(m-1)}$ and $\boldsymbol{\nu}$:
	\begin{equation}
		\boldsymbol{\theta}^{\ast}(\alpha)
		=
		\boldsymbol{\mu}_{\theta} + \big(\boldsymbol{\theta}^{(m-1)}-\boldsymbol{\mu}_{\theta}\big)\cos\alpha + \boldsymbol{\nu}\sin\alpha,
		\qquad \alpha \in [0,2\pi].
		\label{eq:ess_proposal}
	\end{equation}
	The construction is such that $\boldsymbol{\theta}^{\ast}(\alpha)$ is marginally distributed as $\mathcal{N}(\boldsymbol{\mu}_{\theta},\boldsymbol{\Sigma}_{\theta})$ for \emph{every} angle. Two special cases are worth noting. At $\alpha = 0$ we recover the current state, $\boldsymbol{\theta}^{\ast}(0) = \boldsymbol{\theta}^{(m-1)}$; at $\alpha = \pi/2$ we obtain an independent prior draw, $\boldsymbol{\theta}^{\ast}(\pi/2) = \boldsymbol{\mu}_{\theta} + \boldsymbol{\nu}$. The angle therefore interpolates between a null move and a bold, prior-sized move, and it is the only quantity that has to be chosen.
	
	\paragraph{The Slice.} The move along \eqref{eq:ess_proposal} is a rotation of the augmented pair $(\boldsymbol{\theta},\boldsymbol{\nu})$ that leaves its joint Gaussian density unchanged. The prior therefore cancels.\footnote{The update rotates the whitened pair $(\mathbf{z},\boldsymbol{\nu})$ into $(\mathbf{z}\cos\alpha + \boldsymbol{\nu}\sin\alpha,\; \boldsymbol{\nu}\cos\alpha - \mathbf{z}\sin\alpha)$, which preserves $\|\mathbf{z}\|^{2} + \|\boldsymbol{\nu}\|^{2}$. It is this joint invariance, not constancy of the prior along the ellipse, that removes the prior from the acceptance condition \citep{murray2010elliptical}.}  The proposal for $\boldsymbol{\theta}$ is evaluated on the likelihood. We then compare it with:
	\begin{equation}
		\log y
		=
		\log \mathcal{L}(\mathbf{Y}|\boldsymbol{\theta}^{(m-1)},\mathbf{A},\mathbf{B}) + \log \omega,
		\qquad \omega\sim\mathrm{Unif}(0,1).
		\label{eq:ess_slice}
	\end{equation}
	A proposal is accepted whenever its log-likelihood \eqref{eq:lik_theta} lies above this threshold and when it satisfies the restrictions set by the researcher. We discuss ours in the empirical section. The angle in \eqref{eq:ess_proposal} is located by shrinking the bracket rather than by tuning:
	\begin{enumerate}
		\item Draw $\alpha \sim \mathrm{Unif}[0,2\pi)$ and initialize the bracket $[\alpha_{\min}, \alpha_{\max}] = [\alpha - 2\pi,\; \alpha]$.
		\item Form $\boldsymbol{\theta}^{\ast}(\alpha)$ from \eqref{eq:ess_proposal} and evaluate the likelihood \eqref{eq:lik_theta}.
		\item If $\log \mathcal{L}(\mathbf{Y}\mid\boldsymbol{\theta}^{\ast}(\alpha),\mathbf{A},\mathbf{B}) \ge \log y$ and $\boldsymbol{\theta}^{\ast}(\alpha)$ fulfills the restrictions, accept: set $\boldsymbol{\theta}^{(m)} = \boldsymbol{\theta}^{\ast}(\alpha)$ and stop.
		\item Otherwise shrink the bracket towards the current state, setting $\alpha_{\min} = \alpha$ if $\alpha < 0$ and $\alpha_{\max} = \alpha$ otherwise, draw a new $\alpha \sim \mathrm{Unif}(\alpha_{\min}, \alpha_{\max})$, and return to step 2.
	\end{enumerate}
	The bracket always contains $\alpha = 0$, at which the proposal coincides with the current state; the latter is feasible and, because $\log \omega < 0$, strictly above the threshold. The update is rejection-free in the sense that every sweep returns a draw, and it needs no stepsize: the bracket adapts itself. In practice we work in the whitened coordinates $\mathbf{z} = \mathbf{D}^{-1}(\boldsymbol{\theta}-\boldsymbol{\mu}_{\theta})$, with $\mathbf{D} = \boldsymbol{\Sigma}_{\theta}^{1/2}$ diagonal, so that $\boldsymbol{\nu}$ is a standard normal draw; this is equivalent to \eqref{eq:ess_proposal} and numerically more stable. The scheme scales well with the dimension of $\boldsymbol{\theta}$, which matters here because the structural block grows with both the number of countries and the number of variables.
	
	\paragraph{Enforcing the Identified Set.}
	The three classes of restrictions defining $\mathcal{D}(\mathcal{S})$ are enforced at different stages of the sampling update. Zero (equality) restrictions are imposed by construction: the slice sampler operates only on the free elements of $\boldsymbol{\theta}$ while leaving the others constant. Sign restrictions are incorporated into the feasibility check in Step~3. Consequently, any proposal that violates a sign restriction is treated in the same manner as one falling below the slice level, causing the sampling bracket to shrink until a feasible proposal is found\footnote{Imposing the sign restrictions as a hard constraint at this point does not affect the reversibility of the slice sampler with respect to \eqref{eq:post_theta}. The argument of \citet{murray2010elliptical} holds for the Gaussian prior multiplied by an arbitrary non-negative function, and the indicator $\mathbf{1}\{\boldsymbol{\theta}\in\mathcal{D}(\mathcal{S})\}$ is one such function; no continuity of the likelihood is required. The hard sign restriction thus amounts to truncating the Gaussian prior to $\mathcal{D}(\mathcal{S})$ rather than to modifying the transition kernel.}. Finally, magnitude (two-sided inequality) restrictions are enforced through an accept--reject step applied to the normalized draw. A proposal is retained only if all inequality constraints are satisfied; otherwise, the previous draw is preserved\footnote{One can also impose the magnitude restrictions in the ESS step. In our implementation, we only implement sign restrictions in the feasibility check of Step~3 given that the number of strict sign restrictions is smaller, leading to fewer tries of the ESS sampler. Rejections at this step are rare; the sampler records the magnitude acceptance rate for every run.}. Together, these steps ensure that every retained draw belongs to the identified set $\mathcal{D}(\mathcal{S})$. We provide an illustration of the sampler's capabilities in the Online Appendix, Section~B.

	\begin{algorithm}[t]
		\caption{MCMC Sampler for the BSMAR Model}
		\label{alg:mcmc_sampler}
		\begin{algorithmic}[1]
			\State \textbf{Initialize} $\mathbf{A}^{(0)}, \mathbf{B}^{(0)}, \kappa_{\mathbf{A}}^{(0)}, \kappa_{\mathbf{B}}^{(0)}, \boldsymbol{\theta}^{(0)}$
			\State \textbf{Compute} initial structural matrices $\mathbf{B}_r^{(0)}, \mathbf{B}_c^{(0)}$ from $\boldsymbol{\theta}^{(0)}$ and covariance components $\boldsymbol{\Sigma}_r^{(0)} = \mathbf{B}_r^{(0)} {\mathbf{B}_r^{(0)}}', \boldsymbol{\Sigma}_c^{(0)} = \mathbf{B}_c^{(0)} {\mathbf{B}_c^{(0)}}'$
			\For{$m = 1$ \textbf{to} $M$}
			\State \textbf{Step 1: Autoregressive Dynamics}
			\State Compute $\mathbf{K}_{\mathbf{A}}$ and $\widehat{\mathbf{A}}$ using $\mathbf{B}^{(m-1)}$ and $\boldsymbol{\Sigma}_c^{(m-1)}$
			\State Draw $\operatorname{vec}(\mathbf{A}^{(m)}) \sim \mathcal{N}(\operatorname{vec}(\widehat{\mathbf{A}}), \boldsymbol{\Sigma}_r^{(m-1)} \otimes \mathbf{K}_{\mathbf{A}}^{-1})$
			
			\State \textbf{Step 2: Cross-Sectional Dependencies}
			\State Compute $\mathbf{K}_{\mathbf{B}}$ and $\widehat{\mathbf{B}}$ using $\mathbf{A}^{(m)}$ and $\boldsymbol{\Sigma}_r^{(m-1)}$
			\State Draw $\operatorname{vec}(\mathbf{B}^{(m)}) \sim \mathcal{N}(\operatorname{vec}(\widehat{\mathbf{B}}), \boldsymbol{\Sigma}_c^{(m-1)} \otimes \mathbf{K}_{\mathbf{B}}^{-1})$
			
			\State \textbf{Step 3: Structural Parameters (Elliptical Slice Sampling)}
			\State Draw candidate auxiliary direction $\boldsymbol{\nu} \sim \mathcal{N}(\mathbf{0}, \boldsymbol{\Sigma}_{\theta})$
			\State Generate slice log-threshold $\log y = \log \mathcal{L}(\mathbf{Y} \mid \boldsymbol{\theta}^{(m-1)}, \mathbf{A}^{(m)}, \mathbf{B}^{(m)}) + \log \omega$, $\omega \sim \mathrm{Unif}(0,1)$
			\State \textbf{Repeat} until acceptable $\alpha$ is found:
			\State \hspace{1cm} Propose $\boldsymbol{\theta}^{\ast}(\alpha) = \boldsymbol{\mu}_{\theta} + (\boldsymbol{\theta}^{(m-1)}-\boldsymbol{\mu}_{\theta})\cos\alpha + \boldsymbol{\nu}\sin\alpha$
			\State \hspace{1cm} Evaluate log-likelihood $\log \mathcal{L}^{\ast} = \log \mathcal{L}(\mathbf{Y} \mid \boldsymbol{\theta}^{\ast}(\alpha), \mathbf{A}^{(m)}, \mathbf{B}^{(m)})$
			\State \hspace{1cm} Shrink bracket $[\alpha_{\min}, \alpha_{\max}]$ if $\log \mathcal{L}^{\ast} < \log y$ or a sign restriction is violated
			\State Reshape the candidate $\boldsymbol{\theta}^{\ast}(\alpha^\star)$ into $\mathbf{B}_r^{\ast}, \mathbf{B}_c^{\ast}$ and normalize
			\State If the magnitude restrictions hold at $\mathbf{B}_c^{\ast}$, set $\boldsymbol{\theta}^{(m)}, \mathbf{B}_r^{(m)}, \mathbf{B}_c^{(m)}$ to the candidate; otherwise retain $\boldsymbol{\theta}^{(m-1)}, \mathbf{B}_r^{(m-1)}, \mathbf{B}_c^{(m-1)}$
			\State Update $\boldsymbol{\Sigma}_r^{(m)} = \mathbf{B}_r^{(m)} {\mathbf{B}_r^{(m)}}'$ and $\boldsymbol{\Sigma}_c^{(m)} = \mathbf{B}_c^{(m)} {\mathbf{B}_c^{(m)}}'$
			
			\State \textbf{Step 4: Shrinkage Parameters}
			\State Draw $\kappa_{\mathbf{A}}^{(m)}$ and $\kappa_{\mathbf{B}}^{(m)}$ from their $\mathcal{GIG}$ conditional posteriors
			\State Update prior covariance matrices $\mathbf{V}_{\mathbf{A}}$ and $\mathbf{V}_{\mathbf{B}}$
			\EndFor
		\end{algorithmic}
	\end{algorithm}

	\section{Studying High-Dimensional International Spillovers}
	\label{sec:empirical}
	
	There has been long-standing and renewed interest in understanding the drivers of inflation, particularly whether demand or supply shocks have played a dominant role in its recent surge. This question has been addressed using single-country SVAR models identified through sign restrictions \citep[e.g.,][]{giannone_primiceri_2025, bergholt2026inflation, della_chang_jansen_pagliacci_2023}. Spillovers during the pandemic period have been analyzed by \citet{di2023pandemic} in a multi-country, multi-sector New Keynesian model.
	
	We contribute to this literature by providing a multi-country perspective that allows for spillovers across countries. In this section, we apply our proposed framework to a panel of 17 economies. We identify country-specific supply and demand shocks and study their spillovers to other countries over time. We begin by describing the data, then discuss our identification strategy and conclude by presenting our empirical findings.

	\subsection{Data and Specification}
	We use quarterly data for a panel of 17 economies spanning North America, Europe, and the Asia--Pacific region. The panel has the United States (USA) ordered first followed by Australia (AUS), Belgium (BEL), Canada (CAN), China (CHN), Germany (DEU), Spain (ESP), Finland (FIN), France (FRA), United Kingdom (GBR), Italy (ITA), Japan (JPN), South Korea (KOR), Netherlands (NLD), Norway (NOR), Portugal (PRT), and Sweden (SWE). For each country, we observe real gross domestic product (GDP) and the consumer price index (CPI). GDP and CPI are taken from the OECD Quarterly National Accounts and Consumer Price Indices databases, both downloaded from the OECD Data Explorer; the series are listed in the Online Appendix, Section~A.
	
	The estimation sample covers the period from 1998Q1 to 2019Q4. To construct the HDs, we use an out-of-sample evaluation period from 2020Q1 to 2024Q4, encompassing both the COVID-19 pandemic and the subsequent global inflation episode. Prior to estimation, GDP and CPI are transformed into yearly growth rates and standardized. We estimate the model using $p=4$ quarterly lags. The choice of the lag length is standard in SVAR studies using quarterly data.

	\subsection{Identification of Supply and Demand in a Multi-Country Setup}
	\label{sec:emp_id_restrictions}
	
	We now turn to the identification of supply and demand shocks in the multi-country BSMAR model and present the full set of restrictions used in our baseline empirical specification. Recall that from \eqref{eq:BSMAR}  the contemporaneous structural impact matrix is given by the bilinear decomposition $\mathbf{B}_0 = \mathbf{B}_c \otimes \mathbf{B}_r$. With output and inflation as the variables in each country block, this structure allows us to identify country-specific supply and demand shocks while accounting for contemporaneous international spillovers.
	
	Our identifying restrictions are summarized in Table~\ref{tab:emp_restrictions}. We motivate these assumptions as follows. First, the sign restrictions imposed on $\mathbf{B}_r$ follow the standard economic interpretation of supply and demand shocks (see, e.g., \citet{giannone_primiceri_2025}). Both shocks are restricted to be expansionary for economic activity: a positive demand shock raises activity and inflation, whereas a positive (favorable) supply shock raises activity while lowering inflation. The two shocks are therefore distinguished by the sign of the inflation response. Own-country effects are positive, consistent with the expected direction of domestic shocks. This is not a separate restriction but a consequence of the magnitude restriction introduced below, which is stated on the own effect including its sign. For the off-diagonal elements, we assume that international spillovers are likely to operate in the same direction as domestic effects, which we implement using soft sign restrictions (Section~\ref{sec:soft_sign}).\footnote{In the Online Appendix, Section~D, we provide a robustness study of the flexibilization of our prior assumptions. Namely, we relax the prior covariance of the off-diagonal entries. Our findings imply that loosening the prior leads to a reduction in the identification of some spillovers. Nonetheless, we find evidence of demand and supply spillovers between countries, with the United States as an important transmitter.}
	
	To separate country-specific shocks from each other, we impose magnitude restrictions based on the assumption that country-specific shocks have their largest impact on the country in which they originate. The zero restrictions impose belief of the prominent role of the US in the global economy. Namely, we assume that the US is not contemporaneously affected by shocks originating in other countries.

	\begin{table}[!htbp]
		\centering
		\caption{Identifying restrictions in the baseline empirical specification.}
		\label{tab:emp_restrictions}
		\renewcommand{\arraystretch}{1.4}
		\begin{threeparttable}
			\begin{tabular}{@{}p{2.6cm} p{4.9cm} p{6.9cm}@{}}
				\toprule
				\textbf{Family} & \textbf{Restriction} & \textbf{Reading} \\
				\midrule
				Sign, hard &
				$\operatorname{sign}(\mathbf{B}_r) = \left(\begin{smallmatrix} + & + \\ - & + \end{smallmatrix}\right)$ &
				Supply raises output, lowers prices; demand raises both \\
				Soft sign &
				$P(b_c^{j_1j_2}>0) = 0.95$ &
				Spillovers keep the domestic sign \\
				Magnitude &
				$b_c^{j_2j_2} \geq |b_c^{j_1j_2}|$ &
				A shock is largest at home \\
				Zero &
				$b_c^{1j_2}=0$, $j_2 \neq 1$ &
				The US is contemporaneously exogenous \\
				Normalization &
				$b_c^{11}=1$ &
				Fixes the scale \\
				\bottomrule
			\end{tabular}
			\begin{tablenotes}
				\footnotesize
				\item \textit{Note:} $j_2$ denotes the country in which the shock originates and $j_1$ the responding country. The hard sign restrictions are imposed inside the elliptical slice sampler, as a feasibility check that shrinks the angular bracket, and the magnitude restriction by an accept--reject step applied to the normalized draw, the zero restrictions as the exact equality $\mathbf{R}_{0}\boldsymbol{\theta}=\mathbf{0}$, and the soft sign restriction through the Gaussian prior $b_{c,0}^{j_1j_2}=0.5$, $v_{c,0}^{j_1j_2}\approx0.0924$, so that $0.5/\sqrt{v_{c,0}^{j_1j_2}}=z_{0.95}$ and the data can override it; the own effects carry the tight prior $b_{c,0}^{jj}=1$, $v_{c,0}^{jj}=0.1$.
			\end{tablenotes}
		\end{threeparttable}
	\end{table}
	
	Taken together, these restrictions define the identified set $\mathcal{D}(\mathcal{S})$ of the baseline specification and allow us to recover, for each country, its supply and demand shocks as well as the cross-border spillovers between countries.
	
	\subsection{Empirical Results}
	We organize the discussion around four sets of objects: SIRFs, FEVDs, out-of-sample HDs, and connectedness measures. The formulas for the results presented in this Section are described in the Online Appendix, Section~C. Diagnostics on the mixing are presented in the Online Appendix, Section~G.1.4. The posterior traces show no drift, and the autoregressive blocks mix well as shown by the inefficiency factors in every specification. The structural block $\mathbf{B}_c$ is the slowest, as the only one drawn subject to weaker identifying restrictions.

\begin{figure}[H]
		\centering
		\includegraphics[width=0.8\textwidth]{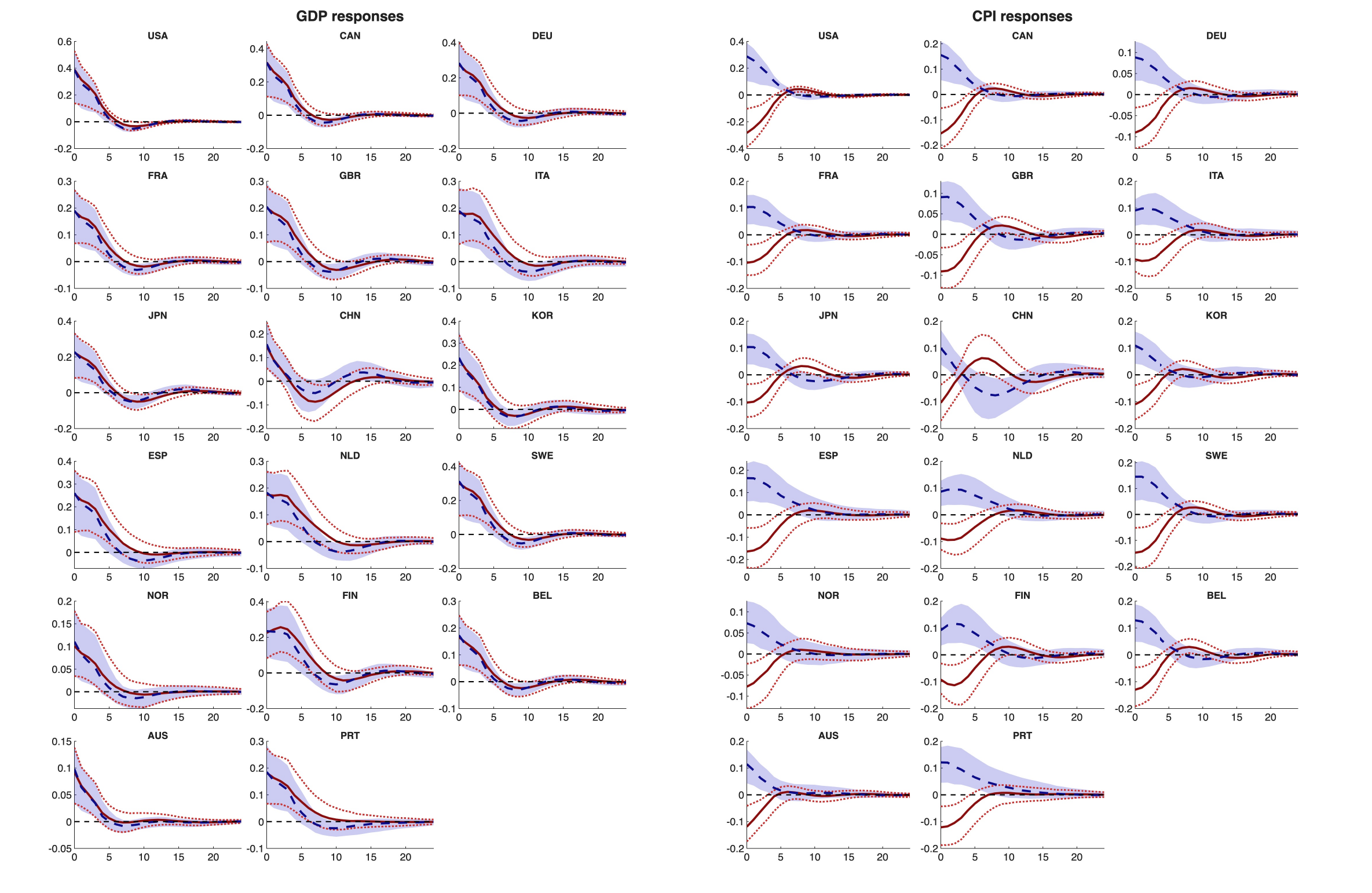}
		\\[0.4em]
		\includegraphics[width=0.7\textwidth]{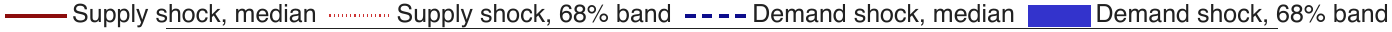}
		\caption{Structural impulse responses of GDP growth (left block) and CPI growth (right block) across the panel to the identified US aggregate supply (AS) shock (solid line, red band) and aggregate demand (AD) shock (dashed line, blue band), baseline specification. Posterior median and 68\% credible band.}
		\label{fig:sirf_usa}
	\end{figure}

	\paragraph{Structural Impulse Responses.}
	Figure~\ref{fig:sirf_usa} reports the structural impulse responses of GDP and CPI growth to the identified US aggregate supply (AS) and aggregate demand (AD) shocks, with the GDP responses in the left block and the CPI responses in the right. The responses to the demand shock show homogeneity in the sign and dynamics, a by-product of the sign restrictions on $\mathbf{B}_r$. More interestingly, the spillovers also maintain the same pattern abroad. That is, this shock leads to an increase in prices and output. The domestic US responses are the largest and the ones with least uncertainty. The responses are largest on impact and die out within about six quarters, after a small overshoot. The responses are mostly similar in shape across countries, while their magnitude and posterior uncertainty differ. Given the dominant role of the US, almost all responses exclude zero from their posterior credible bands. For shocks originating in other countries, we find, as expected, wider posterior bands, since smaller economies play a lesser role in driving the system. Not all responses are as precise as the ones from the US. Whether the responses to other shocks exclude the zero line varies case by case. 
	
	\paragraph{Forecast Error Variance Decompositions.}
	The FEVD decomposes the shock contributions by the region the shock originates from. The results from the sampler for this measure are reported in the Online Appendix in Section~G.1.1. The groups are: domestic, North America, Western Europe, and East Asia and the Pacific at each forecast horizon $h=1,\ldots,24$. The US decomposition is entirely domestic on impact, as the exogeneity restriction requires. At longer horizons foreign shocks, mostly from Western Europe and East Asia, account for a growing share, which is larger for CPI than for GDP. For every other country, shocks originating in Western Europe, the region with the most countries in the panel, account for a larger share of the forecast error variance than domestic shocks. Domestic shocks matter most for Norway, Japan, China and Australia. The shares change little with the horizon, so that the spillovers are already in place on impact. Canada loads on both North America and Western Europe, and the East Asian and Pacific economies retain a visible own-region block alongside their exposure to Europe. The same section reports the decomposition by shock type, which splits each of these blocks into supply and demand. We find that supply and demand shocks contribute roughly equal shares. Supply shocks account for slightly more of the variance of GDP at longer horizons, and demand shocks for slightly more of the variance of CPI. Results from other countries are also presented there.

\begin{figure}[H]
		\centering
		\includegraphics[width=0.7\textwidth]{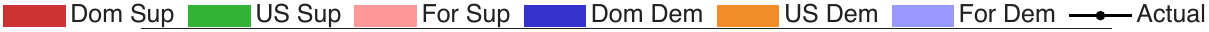}
		
		\vspace{0.4em}
		\includegraphics[width=0.75\textwidth]{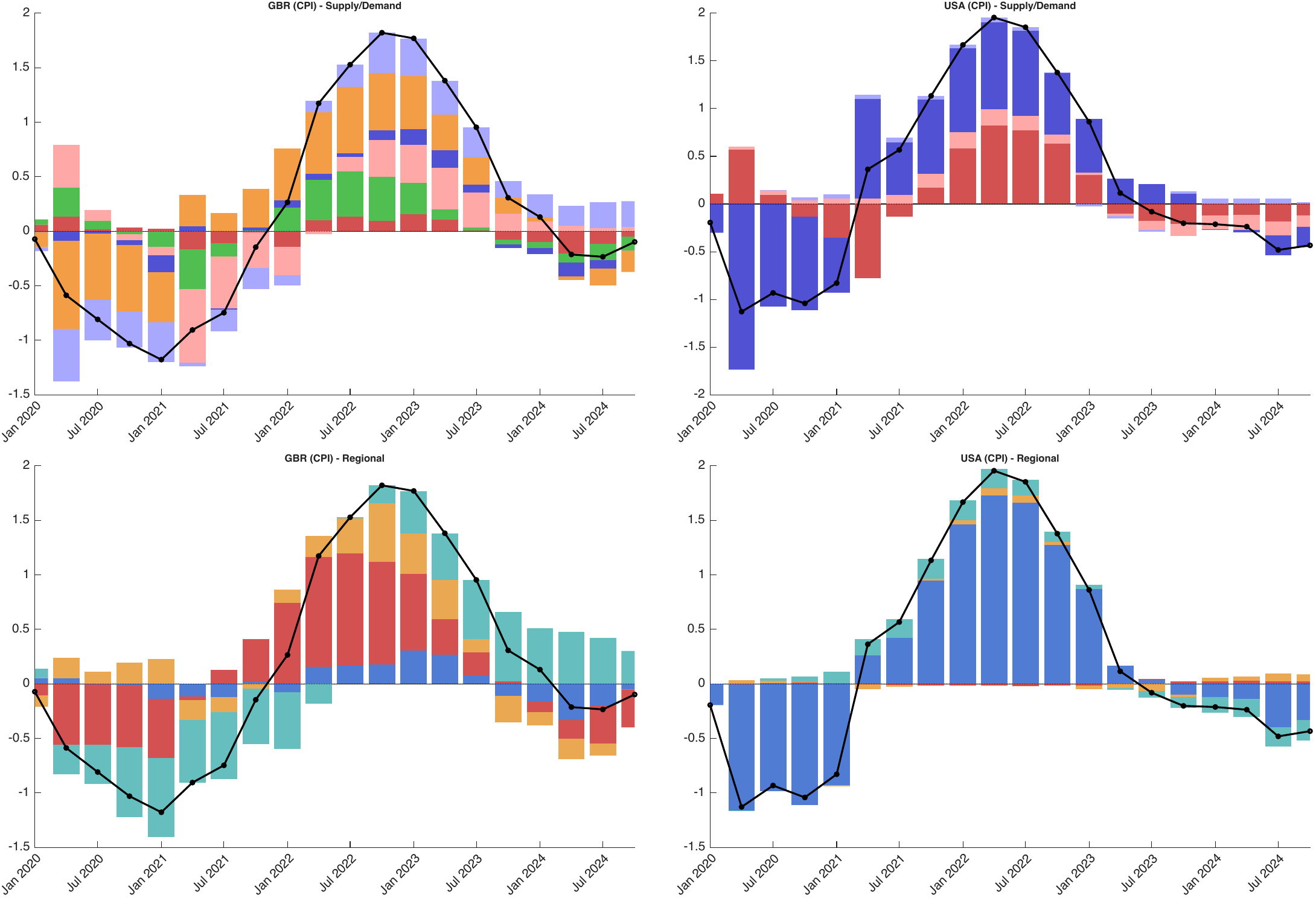}
		
		\vspace{0.4em}
		\includegraphics[width=0.65\textwidth]{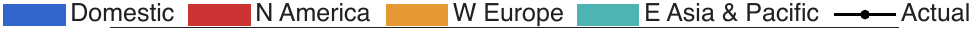}
		\caption{Out-of-sample historical decomposition of CPI for the UK (left column) and the US (right column), baseline specification. Top row: supply/demand decomposition; bottom row: regional decomposition. The supply/demand legend is shown above; the regional legend below.}
		\label{fig:HD_oos_CPI_US_GBR}
	\end{figure}

	\paragraph{Historical Decompositions.}
	We follow \citet{giannone_primiceri_2025} by calculating out-of-sample HDs for the period 2020Q1--2024Q4 to analyze whether the recent increase in inflation was demand or supply driven. Figure~\ref{fig:HD_oos_CPI_US_GBR} reports the decomposition of UK CPI in the left column and US CPI in the right. The figure has four panels: the upper row splits each CPI series into domestic and foreign supply and demand shocks, while the lower row disaggregates the same series by region of origin. In the US, demand shocks clearly dominate the development of CPI during the pandemic. Domestic demand shocks drive both the fall in inflation in 2020 and most of its rise in 2021 and 2022, when domestic supply shocks add a smaller positive contribution. The origin of the shocks is mostly domestic, with minor contributions from Western Europe, East Asia, and the Pacific. The UK, by contrast, shows a more balanced contribution of supply and demand shocks over the COVID period, and almost all of it originates abroad. North American shocks account for most of both the fall in UK inflation in 2020 and 2021 and its rise in 2022, while East Asian shocks pull it down early in the period and push it up after 2023. Placing the two economies side by side reinforces this contrast. US inflation is mainly demand-driven and domestic in origin, whereas UK inflation is driven by foreign shocks, with supply shocks playing a larger role. Overall, and in line with \citet{giannone_primiceri_2025}, these decompositions point to demand shocks as the main explanatory force behind the surge of US inflation during the pandemic. The HDs for the remaining countries and variables are presented in the Online Appendix, Section~G.1.2. In them, the foreign contributions come mostly from Western Europe, the largest group of economies in the panel, with North American and East Asian shocks contributing smaller and similar shares. 
	
\begin{figure}[H]
		\centering
		\includegraphics[width=0.42\textwidth]{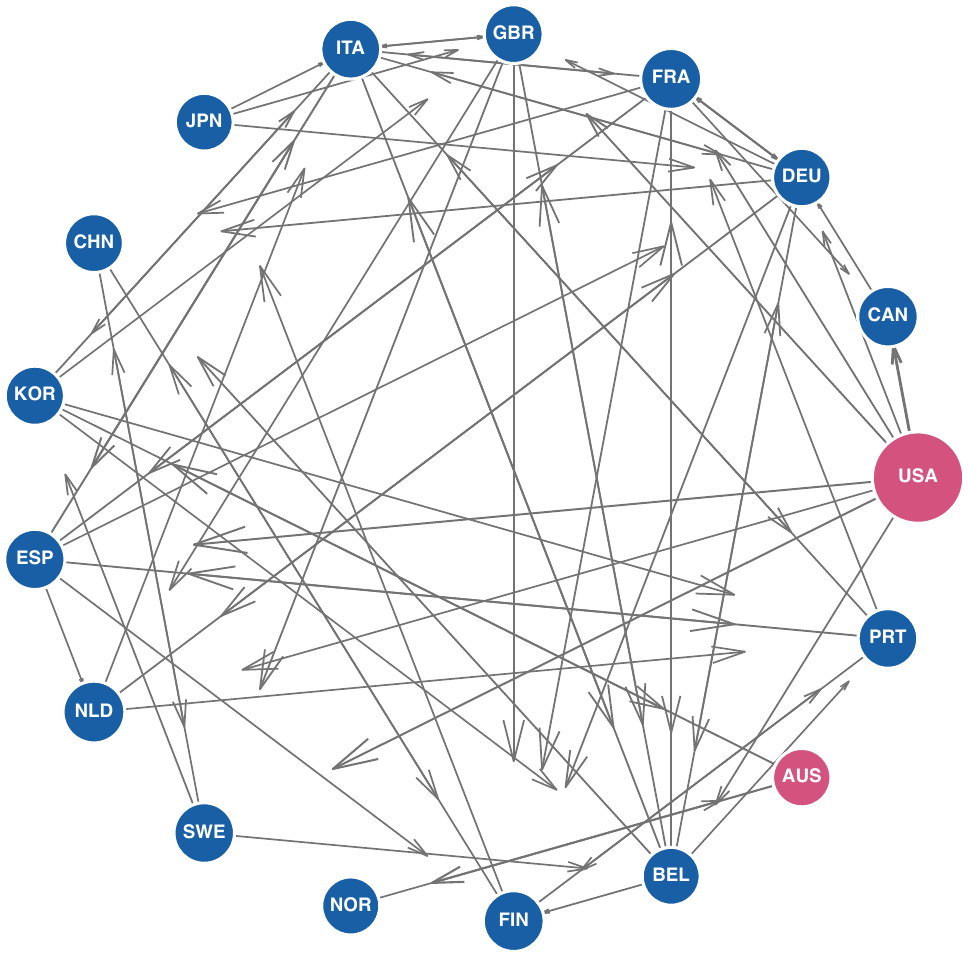}
		
		\vspace{0.4em}
		\includegraphics[width=0.5\textwidth]{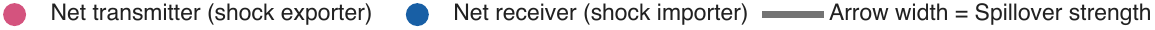}
		\caption{Connectedness network of the spillover effects between countries on impact ($H=1$), from connectedness tables computed for each posterior draw and averaged over the draws.}
		\label{fig:Connectedness_network_firstexo}
	\end{figure}

	\paragraph{Connectedness.}
	Another indicator to summarize the spillover effects is the connectedness measure of \citet{diebold2014network}. Figure~\ref{fig:Connectedness_network_firstexo} presents the network implied by this result, showing the spillovers between countries in both magnitude and direction: the thickness of each edge represents the strength of the directional spillover, while the color of each node indicates whether the country is a net transmitter (red) or a net receiver (blue) and its size the magnitude of that net position. The US is by far the largest net transmitter, with its largest spillover to Canada and visible spillovers to most European economies. Australia is a small net transmitter, while the remaining countries are net receivers. The ranking of spillovers changes with the horizon: short horizons display fewer connections, because the contemporaneous exogeneity of the US binds directly on the impact matrix. We report results for more horizons and under alternative assumptions in the Online Appendix, Sections~G.1.3 and~ G.2.3.\footnote{Another interesting measure of connectedness is presented in \citet{buchwalter2026clustered}. The application of this new measure is left as an extension for future work.}
	
	\FloatBarrier
	\section{Summary}
	\label{sec:summary}
	This paper develops a Bayesian Structural Matrix Autoregression (BSMAR) framework for identifying country-specific macroeconomic shocks and their contemporaneous transmission across countries in large international systems. By exploiting the matrix structure of the data and decomposing the structural impact matrix into variable- and country-specific components, the framework substantially reduces the dimensionality of the model while allowing for rich cross-country spillovers. The proposed Bayesian sampler accommodates a broad class of identification restrictions, providing a flexible framework for structural analysis in high-dimensional systems.
	
	Applying the model to 17 economies, we find a vast number of spillovers between countries. Our results also, partly, validate the findings of the literature by establishing the dominance of demand shocks during the pandemic period. The United States emerges as an important transmitter of shocks, generating sizable spillovers to the rest of the countries in the panel. Compared with the benchmark model in the literature, we find that our model identifies a large number of spillovers under the sign restrictions placed on the behavior of the data, while allowing for violations of these assumptions. Overall, our findings underscore the importance of jointly accounting for country-specific shocks and contemporaneous spillovers when studying the international transmission of macroeconomic disturbances. An interesting direction for future research would be the identification of common shocks that affect all countries at once, e.g., oil or commodity price shocks. This would allow the model to separate global and country-specific shocks. We leave this as a path for future research.

  \bibliography{refs}

@article{chen2021autoregressive,
	title     = {Autoregressive Models for Matrix-Valued Time Series},
	author    = {Chen, Rong and Xiao, Han and Yang, Dan},
	journal   = {Journal of Econometrics},
	volume    = {222},
	number    = {1},
	pages     = {539--560},
	year      = {2021},
	doi       = {10.1016/j.jeconom.2020.07.015}
}

@article{chan2025large,
	title={Large {Bayesian} matrix autoregressions},
	author={Chan, Joshua CC and Qi, Yaling},
	journal={Journal of Econometrics},
	volume={256},
	pages={105955},
	year={2026},
	doi={10.1016/j.jeconom.2025.105955}
}

@article{samadi2025matrix,
	title={On a matrix-valued autoregressive model},
	author={Samadi, S Yaser and Billard, Lynne},
	journal={Journal of Time Series Analysis},
	volume={46},
	number={1},
	pages={3--32},
	year={2025},
	doi={10.1111/jtsa.12748}
}

@article{celani2024matrix,
	title     = {Matrix Autoregressive Models: Generalization and {Bayesian} Estimation},
	author    = {Celani, Alessandro and Pagnottoni, Paolo},
	journal   = {Studies in Nonlinear Dynamics \& Econometrics},
	volume    = {28},
	number    = {2},
	pages     = {227--248},
	year      = {2024},
	doi       = {10.1515/snde-2022-0093}
}

@article{celani2024bayesian,
	title   = {{Bayesian} Variable Selection for Matrix Autoregressive Models},
	author  = {Celani, Alessandro and Pagnottoni, Paolo and Jones, Galin},
	journal = {Statistics and Computing},
	volume  = {34},
	number  = {2},
	pages   = {91},
	year    = {2024},
	doi     = {10.1007/s11222-024-10402-y}
}

@article{bucci2024smooth,
	title   = {A Smooth Transition Autoregressive Model for Matrix-Variate Time Series},
	author  = {Bucci, Andrea},
	journal = {Computational Economics},
	volume  = {65},
	number  = {1},
	pages   = {429--458},
	year    = {2025},
	doi     = {10.1007/s10614-024-10568-7}
}

@misc{chan2024bayesian,
	title={{Bayesian} dynamic factor models for high-dimensional matrix-valued time series},
	author={Chan, Joshua CC and Zhang, Wei},
	year={2024},
	url={https://arxiv.org/abs/2409.08354},
	note = {arXiv preprint 2409.08354 [econ.EM]}
}

@article{forni2001generalized,
	title={The generalized dynamic factor model: representation theory},
	author={Forni, Mario and Lippi, Marco},
	journal={Econometric Theory},
	volume={17},
	number={6},
	pages={1113--1141},
	year={2001}
}

@techreport{stock2005implications,
	title={Implications of dynamic factor models for {VAR} analysis},
	author={Stock, James H and Watson, Mark W},
	institution={National Bureau of Economic Research},
	type={NBER Working Paper},
	number={11467},
	year={2005},
	doi={10.3386/w11467}
}

@article{charnavoki2014effects,
	title={The Effects of Global Shocks on Small Commodity-Exporting Economies: Lessons from {C}anada},
	author={Charnavoki, Valery and Dolado, Juan J.},
	journal={American Economic Journal: Macroeconomics},
	doi={10.1257/mac.6.2.207},
	volume={6},
	number={2},
	pages={207--237},
	year={2014}
}

@article{mumtaz2009transmission,
	title={The transmission of international shocks: a factor-augmented {VAR} approach},
	author={Mumtaz, Haroon and Surico, Paolo},
	journal={Journal of Money, Credit and Banking},
	volume={41},
	number={s1},
	pages={71--100},
	year={2009},
	doi={10.1111/j.1538-4616.2008.00199.x}
}

@article{pesaran2004modeling,
	title={Modeling regional interdependencies using a global error-correcting macroeconometric model},
	author={Pesaran, M Hashem and Schuermann, Til and Weiner, Scott M},
	journal={Journal of Business \& Economic Statistics},
	volume={22},
	number={2},
	pages={129--162},
	year={2004}
}

@article{dees2007exploring,
	title   = {Exploring the International Linkages of the Euro Area: A Global {VAR} Analysis},
	author  = {Dees, Stephane and di Mauro, Filippo and Pesaran, M. Hashem and Smith, L. Vanessa},
	journal = {Journal of Applied Econometrics},
	volume  = {22},
	number  = {1},
	pages   = {1--38},
	year    = {2007},
	doi     = {10.1002/jae.932}
}

@article{feldkircher2016international,
	title={The international transmission of {US} shocks—Evidence from {Bayesian} global vector autoregressions},
	author={Feldkircher, Martin and Huber, Florian},
	journal={European Economic Review},
	volume={81},
	pages={167--188},
	year={2016},
	doi={10.1016/j.euroecorev.2015.01.009}
}

@article{crespo2019spillovers,
	title={Spillovers from {US} monetary policy: evidence from a time varying parameter global vector auto-regressive model},
	author={Crespo Cuaresma, Jesus and Doppelhofer, Gernot and Feldkircher, Martin and Huber, Florian},
	journal={Journal of the Royal Statistical Society Series A: Statistics in Society},
	volume={182},
	number={3},
	pages={831--861},
	year={2019},
	doi={10.1111/rssa.12439}
}

@article{pfarrhofer2025high,
	title={High-frequency and heteroskedasticity identification in multicountry models: Revisiting spillovers of monetary shocks},
	author={Pfarrhofer, Michael and Stelzer, Anna},
	journal={Macroeconomic Dynamics},
	volume={29},
	pages={e122},
	year={2025},
	doi={10.1017/S136510052510031X}
}

@article{camehl2026explains,
	title={What Explains International Interest Rate Co-Movement?},
	author={Camehl, Annika and von Schweinitz, Gregor},
	journal={Journal of Applied Econometrics},
	volume={41},
	number={4},
	pages={343--359},
	year={2026},
	doi={10.1002/jae.70044}
}

@techreport{di2023pandemic,
	title={Pandemic-era inflation drivers and global spillovers},
	author={Di Giovanni, Julian and Kalemli-{\"O}zcan, {\c S}ebnem and Silva, Alvaro and Yildirim, Muhammed A.},
	year={2023},
	institution={National Bureau of Economic Research},
	type={NBER Working Paper},
	number={31887},
	doi={10.3386/w31887}
}

@article{hou2024large,
	title={Large {Bayesian} {SVARs} with linear restrictions},
	author={Hou, Chenghan},
	journal={Journal of Econometrics},
	volume={244},
	number={1},
	pages={105850},
	year={2024}
}

@article{sims1980macroeconomics,
	title={Macroeconomics and reality},
	author={Sims, Christopher A},
	journal={Econometrica},
	volume={48},
	number={1},
	pages={1--48},
	year={1980},
	doi={10.2307/1912017}
}

@article{blanchard_quah_1989,
	author  = {Blanchard, Olivier J. and Quah, Danny},
	title   = {The Dynamic Effects of Aggregate Demand and Supply Disturbances},
	journal = {American Economic Review},
	year    = {1989},
	volume  = {79},
	number  = {4},
	pages   = {655--673}
}

@article{rubio_ramirez_waggoner_zha_2010,
	author  = {Rubio-Ram{\'\i}rez, Juan F. and Waggoner, Daniel F. and Zha, Tao},
	title   = {Structural Vector Autoregressions: Theory of Identification and Algorithms for Inference},
	journal = {Review of Economic Studies},
	year    = {2010},
	volume  = {77},
	number  = {2},
	pages   = {665--696}
}

@article{arias_rubio_ramirez_waggoner_2018,
	author  = {Arias, Jonas E. and Rubio-Ram{\'\i}rez, Juan F. and Waggoner, Daniel F.},
	title   = {Inference Based on Structural Vector Autoregressions Identified with Sign and Zero Restrictions: Theory and Applications},
	journal = {Econometrica},
	year    = {2018},
	volume  = {86},
	number  = {2},
	pages   = {685--720},
	doi     = {10.3982/ECTA14468}
}

@techreport{arias_rubio_ramirez_shin_2025,
	author      = {Arias, Jonas E. and Rubio-Ram{\'\i}rez, Juan F. and Shin, Minchul},
	title       = {Large {SVAR}s},
	institution = {Federal Reserve Bank of Philadelphia},
	type        = {Working Paper},
	year        = {2025}
}

@article{baumeister2015sign,
	title={Sign restrictions, structural vector autoregressions, and useful prior information},
	author={Baumeister, Christiane and Hamilton, James D},
	journal={Econometrica},
	volume={83},
	number={5},
	pages={1963--1999},
	year={2015}
}

@techreport{read2025fast,
	title={Fast Posterior Sampling in Tightly Identified {SVAR}s Using `Soft' Sign Restrictions},
	author={Read, Matthew and Zhu, Dan},
	institution={Reserve Bank of Australia},
	type={Research Discussion Paper},
	number={2025-03},
	year={2025},
	doi={10.47688/rdp2025-03}
}

@inproceedings{murray2010elliptical,
	title   = {Elliptical slice sampling},
	author  = {Murray, Iain and Adams, Ryan P. and MacKay, David J. C.},
	booktitle = {Proceedings of the Thirteenth International Conference on Artificial Intelligence and Statistics},
	year    = {2010},
	volume  = {9},
	pages   = {541--548},
	editor  = {Teh, Yee Whye and Titterington, Mike},
	series  = {Proceedings of Machine Learning Research},
	publisher = {PMLR},
	url     = {https://proceedings.mlr.press/v9/murray10a.html}
}

@article{Vrugt2016_DREAM,
	title        = {Markov chain {Monte Carlo} simulation using the {DREAM} software package: Theory, concepts, and {MATLAB} implementation},
	author       = {J.~A. Vrugt},
	journal      = {Environmental Modelling \& Software},
	year         = {2016},
	volume       = {75},
	pages        = {273--316},
	doi          = {10.1016/j.envsoft.2015.08.013}
}

@article{diebold2014network,
	title={On the network topology of variance decompositions: Measuring the connectedness of financial firms},
	author={Diebold, Francis X and Y{\i}lmaz, Kamil},
	journal={Journal of Econometrics},
	volume={182},
	number={1},
	pages={119--134},
	year={2014}
}

@article{buchwalter2026clustered,
	title={Clustered network connectedness: A new measurement framework with application to global equity markets},
	author={Buchwalter, Bastien and Diebold, Francis X and Yilmaz, Kamil},
	journal={Journal of Econometrics},
	pages={106243},
	year={2026},
	publisher={Elsevier}
}

@article{litterman1986,
	author  = {Litterman, Robert B.},
	title   = {Forecasting with {B}ayesian Vector Autoregressions: Five Years of Experience},
	journal = {Journal of Business \& Economic Statistics},
	volume  = {4},
	number  = {1},
	pages   = {25--38},
	year    = {1986}
}

@article{giannone_lenza_primiceri_2015,
	author  = {Giannone, Domenico and Lenza, Michele and Primiceri, Giorgio E.},
	title   = {Prior Selection for Vector Autoregressions},
	journal = {The Review of Economics and Statistics},
	volume  = {97},
	number  = {2},
	pages   = {436--451},
	year    = {2015}
}

@unpublished{giannone_primiceri_2025,
	author      = {Giannone, Domenico and Primiceri, Giorgio E.},
	title       = {Demand-Driven Inflation},
	year        = {2025},
	month       = aug,
	note        = {Working paper},
	url         = {https://faculty.wcas.northwestern.edu/gep575/DemandDrivenInflation_4-3.pdf}
}

@article{della_chang_jansen_pagliacci_2023,
	author  = {Chang, Jui-Chuan Della and Jansen, Dennis W. and Pagliacci, Carolina},
	title   = {Inflation and Real {GDP} Growth in the {U.S.}---Demand or Supply Driven?},
	journal = {Economics Letters},
	year    = {2023},
	volume  = {231},
	pages   = {111274},
	doi     = {10.1016/j.econlet.2023.111274}
}

@article{bergholt2026inflation,
	author  = {Bergholt, Drago and Canova, Fabio and Furlanetto, Francesco and Maffei-Faccioli, Nicol{\`o} and Ulvedal, P{\aa}l},
	title   = {What Drives the Recent Surge in Inflation? {The} Historical Decomposition Roller Coaster},
	journal = {American Economic Journal: Macroeconomics},
	year    = {2026},
	volume  = {18},
	number  = {4},
	pages   = {34--63},
	doi     = {10.1257/mac.20240209}
}

@book{kilian2017structural,
	title={Structural vector autoregressive analysis},
	author={Kilian, Lutz and L{\"u}tkepohl, Helmut},
	year={2017},
	publisher={Cambridge University Press}
}

@article{devroye2014random,
	author  = {Devroye, Luc},
	title   = {Random Variate Generation for the Generalized Inverse Gaussian Distribution},
	journal = {Statistics and Computing},
	year    = {2014},
	volume  = {24},
	number  = {2},
	pages   = {239--246}
}

@article{kong2026matrix,
	title   = {Matrix Quantile Factor Model},
	author  = {Kong, Xinbing and Liu, Yongxin and Yu, Long and Zhao, Peng},
	journal = {Journal of Business \& Economic Statistics},
	year    = {2026},
	note    = {Forthcoming},
	doi     = {10.1080/07350015.2026.2719893}
}

@article{ye2026multi,
	title   = {Multi-Matrix Autoregressive Models with an Application to Multi-Modal Network},
	author  = {Ye, Shiqi and Zheng, Tingguo and Xiao, Han and Hong, Yongmiao},
	journal = {Journal of Business \& Economic Statistics},
	year    = {2026},
	note    = {Forthcoming},
	doi     = {10.1080/07350015.2026.2682541}
}

@book{rue_held_2005,
	author    = {Rue, H{\aa}vard and Held, Leonhard},
	title     = {Gaussian Markov Random Fields: Theory and Applications},
	publisher = {Chapman \& Hall/CRC},
	address   = {Boca Raton},
	year      = {2005}
}

\end{document}